\documentclass[11pt]{article}
\usepackage{times}
\usepackage{latexsym}
\usepackage{epsfig,enumerate,amsmath,amsfonts,amssymb,setspace}
\usepackage{indentfirst}
\usepackage{wrapfig,lipsum}
\usepackage{epsfig}
\usepackage{graphicx}
\usepackage{amsmath}
\usepackage{amsthm}
\usepackage{tikz}

\usetikzlibrary{positioning}
\usetikzlibrary{shapes.geometric}
\usepackage{algorithm}
\usepackage{algorithmic}

\usepackage{url}
\usepackage[colorlinks=true,
            linkcolor=blue,
            citecolor=blue,
            urlcolor=blue]{hyperref}

\usepackage{cleveref}
\providecommand{\keywords}[1]{\textbf{Keywords:} #1}
\usepackage[T1]{fontenc}
\usepackage[utf8]{inputenc}
\usepackage{authblk}
\usepackage[algo2e,ruled,vlined]{algorithm2e}
\newtheorem{definition}{Definition}[section]
\newtheorem{theorem}{Theorem}[section]

\newtheorem{corollary}{Corollary}[section]

\newtheorem{observation}{Observation}
\theoremstyle{remark}
\newtheorem{remark}{Remark}[section]
\newtheorem{proposition}{Proposition}
\newcommand{\modo}{{\rm mod}}
\newcommand{\2}{ \vspace{0.2cm} }

\newcommand{\vt}[1]{{\color{black}#1}}

\title{Odd Chromatic Number of Graph Classes: Bounds and Structural Results}

\author[1]{Musarof Hossain \thanks{d24174@students.iitmandi.ac.in}}
\author[1]{Sandeep Kumar \thanks{d24176@students.iitmandi.ac.in}}
\author[1]{Vikash Tripathi \thanks{vikash@iitmandi.ac.in}}
\author[1]{Mirza Galib Anwarul Husain Baig \thanks{mirza@iitmandi.ac.in}}
\affil[1]{Indian Institute of Technology Mandi, Himachal Pradesh, India}

\begin{document}
\maketitle

\begin{abstract}
    A proper coloring of a graph $G$  is called an \emph{odd coloring} if, for every non-isolated vertex $v\in V(G)$, there exists a color that occurs an odd number of times in the open neighborhood of $v$. For a graph $G$, the minimum number of colors required for such a coloring is called the {\em odd chromatic number}, denoted by $\chi_{o}(G)$. The odd chromatic number has been well-studied from both combinatorial and algorithmic perspectives, and several bounds on $\chi_o(G)$ are known. In this paper, we establish tight bounds on $\chi_o(G)$ for several graph classes.
In particular, we obtain the following results:
\begin{itemize}
\item For every block-cactus graph $G$, we prove that $\chi_o(G) \leq \max \{5,\omega(G)+1\}$, where $\omega(G)$ denotes the clique number of $G$.

\item For block graphs and proper interval graphs, we establish the bounds $\omega(G) \leq \chi_o(G) \leq \omega(G)+1$. 

\item For a bipartite circle graph, we prove that $\chi_o(G) \leq 6$. This settles an open question asked by Davis et al., 2026.

\item For every biconvex bipartite graph $G$, we prove that $\chi_o(G) \leq 5$.

\item For $C_4$-free cographs $G$, we establish the bound $\chi_o(G) \leq \omega(G)+1$.

\end{itemize}
Moreover, we characterize block graphs, proper interval graphs, and cactus graphs based on their odd chromatic number.
\end{abstract}


\noindent
\keywords{Odd Chromatic Number;  Block-Cactus Graphs;  Block Graphs;  Proper Interval Graphs;  Bipartite Circle Graphs;  Biconvex Bipartite Graphs;   Cographs.}


\section{Introduction}

Throughout the paper, we have considered simple, connected, and finite graphs. We use $[n]$ to denote the set $\{1,2, \ldots,n\}$. Coloring in graphs has been one of the fundamental problems in graph theory since its inception. Among several graph coloring parameters, the chromatic number of graphs is an extensively studied parameter in graph theory and has enormous applications in the real world; see~\cite{coloring_book_1,coloring_book_2}. A proper $k$-coloring of a graph $G$ is a map $f:V(G)\rightarrow [k]$ such that if $uv \in E(G)$ then $f(u) \neq f(v)$. The minimum $k$ for which there exists a proper coloring $f:V(G)\rightarrow [k]$ is called {\em chromatic number} of $G$, symbolized as~$\chi(G)$. 


Motivated by various real-world applications, numerous variants and generalizations of proper coloring have been introduced~\cite{caro2023remarks,fabrici2023proper,CD_coloring, dominator_coloring}. Among the various extensions of proper graph coloring, proper conflict-free coloring (PCFC) is a well-studied variant~\cite{caro2023remarks,fabrici2023proper}, in which every vertex $v$ is required to have at least one color that occurs exactly once in its open neighborhood of $v$. Along similar lines, Petruševeski and Skrekovski \cite{petruvsevski2022colorings} introduced odd coloring as a further generalization of PCFC. 

A proper $k$-coloring of a graph $G$ is called an {\em odd $k$-coloring} if, for every non-isolated vertex $v\in V(G)$, there exists a color that occur an odd number of times in the open neighborhood of $v$. The odd chromatic number of $G$, denoted by $\chi_o(G)$, is the minimum number of colors for which $G$ admits an odd coloring. In the associated decision problem called Odd $k$-Coloring, given a graph $G$ and integer $k>0$, we need to answer whether $G$ admits an odd coloring using at most $k$-colors or not. It is known that the Odd $k$-Coloring is NP-hard for bipartite graphs \cite{ahn2025proper}. Later, Patar and Reddy \cite{patar2026odd} strengthened the hardness results by showing that the problem remains hard even for perfect elimination bipartite graphs and star-convex bipartite graphs.
Bhyravarapu et al. \cite{bhyravarapu2025parameterized} investigated the parameterized complexity of odd coloring and showed that Odd $k$-Coloring is fixed-parameter tractable when parameterized by distance to cluster, distance to co-cluster, and neighborhood diversity. In addition, they obtained polynomial-time algorithms for Odd $k$-Coloring on cographs and split graphs. Patar and Reddy \cite{patar2026odd} studied the problem for chain graphs and convex bipartite graphs. 

In the context of combinatorial bounds, Daniel W. Cranston \cite{cranston2022odd} determined the odd chromatic number of cycles and trees. In particular, they prove that for a cycle on $n$ vertices $C_n$, $\chi_o(C_n) \le 5$ and for a path on $n$ vertices $P_n$, $\chi_o(P_n) \le 3$. Caro, Petruševski, and Škrekovski \cite{CARO2022392} prove that $\chi_{o}(G) \le 9$ when $G$ is a connected planar graph. They also conjectured that for a planar graph $G$, $\chi_{o}(G) \le 5$. Later, in \cite{CARO2022392}, the authors reduced the bound to $8$ for planar graphs. Several other bounds have also been established; see~\cite{cranston2023note,liu2026odd,cho2023odd,kashima2024odd,pradhan2026remark}.
Pang, Miao, and Fan \cite{pang2026odd} studied the odd chromatic number in terms of the maximum degree of a graph and conjectured that every connected graph $G$ with maximum degree $\Delta(G)\ge3$ satisfies $\chi_o(G)\le 2\Delta(G)-1$.



\medskip
\noindent\textbf{Our contributions.}
In this paper, we investigate the odd chromatic number for several well studied graph classes. Our study focuses block-cactus graphs, block graphs, proper interval graphs, biconvex bipartite graphs, bipartite circle graphs, bipartite distance-hereditary graphs, and $C_4$-free cographs. The aim is to use the structures of these graph classes to establish tight lower and upper bounds. We identify that the odd chromatic number is very close to the clique number. It is important to remark that, for some graph classes, the bounds that we establish settles the conjecture posed by Pang et al. in \cite{pang2026odd}.



We first consider block cactus graphs and establish the bound
$\chi_o(G)\leq \max\{5,\omega(G)+1\}$,
where $\omega(G)$ denotes the clique number of $G$. Next, we show that if $G$ is a block graph or a proper interval graph then $\omega(G) \le \chi_{o}(G) \le \omega(G)+1$ and characterize the graphs attaining both lower and upper bounds. Moreover, we design an efficient algorithm for odd coloring of proper interval graphs with minimum colors. 

Davies et al. \cite{Davies2025OddColoring} posed Problem 5.2, asking whether every bipartite circle graph has odd chromatic number at most four. We answer this question in the negative by constructing a connected bipartite distance-hereditary graph, which is also a bipartite circle graph, with odd chromatic number five. Furthermore, we prove that every bipartite circle graph admits a proper odd coloring using at most six colors and demonstrate that this bound is sharp. Consequently, our result improves the previously best-known upper bound of 98, established in Corollary 1.6 of \cite{Davies2025OddColoring}, to the sharp upper bound of six.

Next, we examine connected biconvex bipartite graphs and show that every such graph admits an odd coloring using at most five colors. The proof utilizes the multichain ordering of biconvex graphs. We have established the sharpness of this bound for bipartite biconvex graphs. For a cograph $G$, it is known that $\chi_{o}(G) \le \omega(G)+2$. We observed that this particular bound is indeed tight for cographs. For example, for $C_4$, which is a cograph, $\chi_{o}(G) = 4 = \omega(G)+2$. We analyze the bound on cograph in the absence of $C_4$ (that is, $C_4$-free cographs), and we improve this bound for $C_4$-free cographs. In particular, for a $C_4$-free cographs we show that $\chi_{o}(G)\le \omega(G)+1$ and prove that this bound is indeed tight. As a result, we settle the conjecture posed by Pang et al. in \cite{pang2026odd} for block-cactus graphs, block graphs, proper interval graphs, biconvex bipartite graphs and $C_4$-free cographs.





\section{Preliminaries}


For a graph $G=(V,E)$, we denote its vertex set and edge set by $V(G)$ and $E(G)$, respectively. For vertices $u,v\in V(G)$, an edge joining $u$ and $v$ is denoted by $uv$. For a subset $S\subseteq V(G)$, we denote by $G[S]$ the subgraph of $G$ induced by $S$. For a vertex $v\in V(G)$, the open neighborhood of $v$ is defined by $N_G(v):=\{u\in V(G)\mid uv\in E(G)\}$, and its closed neighborhood is defined by $N_G[v]:=N_G(v)\cup\{v\}$. When the underlying graph is clear from the context, we simply write $N(v)$ and $N[v]$, respectively. A vertex $v\in V(G)$ is called a \emph{universal vertex} if it is adjacent to every other vertex of $G$. The degree of a vertex, denoted as ${\rm deg}_{G}(v)$, is the cardinality of $N_{G}(v)$. A bridge (cut-edge) in a graph $G$ is an edge whose removal increases the number of connected components in the graph. 

We next introduce the terminology associated with colorings that will be used throughout the paper. Let $f:V(G)\rightarrow[k]$ be a proper coloring of $G$. For each $i\in[k]$, we denote by $f^{-1}(i)=\{u\in V(G)\mid f(u)=i\}$ the set of vertices assigned color $i$ under $f$. A vertex $v\in V(G)$ is said to have $c\in[k]$ as an \emph{odd color} with respect to $f$ if an odd number of its neighbors are assigned color $c$. Equivalently, $c$ is an odd color of $v$ with respect to $f$ if $|f^{-1}(c)\cap N_G(v)|\equiv 1\pmod{2}$.

In a graph $G$, a set $S\subseteq V(G)$ is a {\em clique} if for any two distinct vertices $u,v \in S$, $uv \in E(G)$. The clique size is the maximum cardinality of a clique in $G$. We use $C_n$ and $P_n$ to denote a cycle and a path on $n$-vertices, respectively. 
A vertex $v\in V(G)$ is called a \emph{cut vertex} if $G[V(G)\setminus\{v\}]$ is disconnected. For a graph $G$, a maximal induced subgraph of $G$ without a cut vertex is called a block of $G$. We use $B, B_1, B_2, \ldots$ to denote blocks of a graph $G$. By the definition of blocks, we note that if $B_1$ and $B_2$ are two blocks, then $|V(B_1) \cap V(B_2)| \le 1$.  Thus, each block is either a maximal $2$-connected subgraph, an edge, or an isolated vertex. A \emph{block graph} is a graph in which every block is a clique, whereas a \emph{block-cactus graph} is a graph in which every block is either a clique or a cycle, and a cactus graph is a connected graph in which every block is either an edge or a cycle. A Chain Graph is a bipartite graph $G=(V_1,V_2,E)$ in which the neighborhoods of the vertices in one partite set are linearly ordered by inclusion; that is, for any $u,v\in V_1$, either $N(u)\subseteq N(v)$ or $N(v)\subseteq N(u)$.

Finally, we recall the notion of perfect graphs. For a graph $G$, let $\chi(G)$ denote its chromatic number and $\omega(G)$ its clique number. A graph $G$ is called \emph{perfect} if, for every induced subgraph $H$ of $G$, $\chi(H)=\omega(H).$ In particular, every block graph, every proper interval graph, and every $C_4$-free cograph is perfect.




\section{Block-Cactus Graphs}
In this section, we study the odd chromatic number in a block-cactus graph. Recall that if $G$ is a block-cactus graph, then $G$ has a block $B$ such that $B$ has exactly one cut vertex and such a block is called an {\em end block}. Before proving the main result, we state the following observations.  

\begin{observation}
    If $G$ is a complete graph, then $\chi_o(G)=\omega(G)=|V(G)|.$
\end{observation}
\begin{observation}
\label{obs:oddchromaticnumber_2}
    For any general graph $G$, $\chi_o(G)=2$ if and only if $G$ is an odd-degree graph, and every cycle in $G$ has an even length.
\end{observation}

\begin{observation}
\label{obs:odd_chromatic_number_for_a_clique}
Let $B$ be a clique block of a block-cactus graph $G$, and let $v\in V(G)$ be a non-cut vertex; that is, $v$ belongs to exactly one block $B$ of $G$. Then there exists a color that occurs an odd number of times in $N(v)$ under any proper coloring of $G$.
\end{observation}

The next theorem establishes the bound on the odd chromatic number of block-cactus graphs.




\begin{theorem}
\label{thm:block_cactus_graph}
    For a block-cactus graph $G$, $2 \le \chi_{o}(G)\le \max\{5,\omega(G)+1\}$.
\end{theorem}
\begin{proof}
    Let $G$ be a block-cactus graph and $M=\max\{5,\omega(G)+1\}$. The lower bound is obvious from the fact that for any connected graph $H$ having at least two vertices, $\chi(H) \ge 2$ and $\chi_o(H) \ge \chi(H)$. To establish the upper bound, we use induction on the number of blocks in the block-cactus graph $G$.

    Suppose $G$ has only one block. In this case, $G$ is either a cycle or a complete graph. If $G$ is a cycle then $\chi_o(G) \le 5$ while if $G$ is a complete graph on $n$ vertices then $\chi_o(G)= n =\omega(G)$. This establishes the base case. Next, we assume that the result holds for any block-cactus graph having at most $r-1$ blocks for $r\ge 2$.

    Now consider a block-cactus graph $ G$ having $r$ blocks. Recall that any block-cactus graph $G$ has at least two end blocks. Let $B$ be an end block with cut vertex $v$.
    Notably, $B$ is a clique or a cycle. Let $S=(V(G)-V(B))\cup \{v\}$ and $H= G[S]$. Note that the graph $H$ is a block-cactus graph having $r-1$ blocks. Therefore, by the inductive hypothesis, $H$ admits an odd coloring $f$ using at most $\max\{\omega(H)+1, 5\}$ colors. Since $\omega(H) \le \omega(G)$, $\max\{\omega(H)+1, 5\} \le M$. Thus, $H$ admits an odd coloring using at most $M$ colors. As $f$ is an odd coloring of the graph $H$ using at most $M$ colors, there exists a color $i \in [M]$ such that $|f^{-1}(i) \cap N(v)|$ is odd. Also, assume that $f(v) = j$. Next, depending on whether $B$ is a complete graph or a cycle, we extend the odd coloring $f$ of $H$ to an odd coloring $f'$ of $G$ such that $f'$ uses at most $M$ colors.

    \medskip 
    \noindent \textbf{Case 1: $B$ is a clique.} 

    In this case, first, we note that $|V(B)| \le \omega(G)$ and no vertex of $V(B) \setminus \{v\}$ has a neighbor in $V(G) \setminus V(B)$. Thus, we extend the coloring $f$ of $H$ to an odd coloring of $G$ as follows: $f'(u)=f(u)$ for each vertex $u \in V(G) \setminus (V(B) \setminus \{v\})$. Assign all vertices in $V(B) \setminus \{v\} $ a distinct color from the set $[M] \setminus \{i,j\}$. Now, we claim that the coloring $f'$ is an odd coloring of $G$. Noting the fact that no vertex of $V(B) \setminus \{v\}$ has a neighbor in $V(G) \setminus V(B)$ and $f$ is a odd coloring of $H$, for any pair of vertices $u,w \in V(G) \setminus V(B)$, if $uw \in E(G)$, $f(u) \neq f(w)$. Also, for each vertex $u \in V(G) \setminus V(B)$, there exists a color $k \in [M]$ such that $|f'^{-1}(k) \cap N_G(u)|$ is odd. 
    
    Next, note that no vertex in $V(B) \setminus \{v\}$ is assigned color $i$ or $j$. Thus $|f'^{-1}(i) \cap N_G(v)|$ remains odd. Also, for each vertex $u \in V(B) \setminus \{v\}$, $|f'^{-1}(j) \cap N_G(u)| = 1$. Precisely, $v$ is the only vertex colored $j$ in the neighborhood of each vertex $u \in V(B) \setminus \{v\}$. This concludes that if $B$ is a clique then the extended coloring $f'$ is an odd coloring of $G$ that uses at most $M$ colors.
    

    
    \medskip 
    \noindent \textbf{Case 2: $B$ is a cycle.}  
    
    Let $B=C_t$ for some $t \ge 3$. Note that $M \ge 5$. Since $B$ is a cycle in $G$, $G[V(B) \setminus \{v\}]$ is path on $t-1$ vertices. 
    Similar to the previous case, we note that no vertex in $V(B) \setminus \{v\}$ has a neighbor in $V(G) \setminus V(B)$. Thus, we extend the odd coloring $f$ of $H$ to an odd coloring $f'$ of $G$ as follows: $f'(u)=f(u)$ for each vertex $u \in V(G) \setminus (V(B) \setminus \{v\})$. Assign colors to the set of vertices in $V(B) \setminus \{v\}$ from the set of colors $M \setminus \{i,j\}$ such that $f'$ restricted to the graph $G[V(B)\setminus \{v\}]$ is an odd coloring. This is indeed possible due to the fact that $|M \setminus \{i,j\}| \ge 3$ and the odd chromatic number of any path is at most $3$. Consequently, $f'$ is an odd coloring of $G$ that uses at most $M$ colors.
    


This completes the proof of the theorem.
\end{proof}

Since in a cactus graph there is no clique of size more than $3$, we have the following bound on cactus graphs.

\begin{corollary}
    \label{cor:cactus_graph}
         Let $G$ be a cactus graph. Then $2 \le \chi_{o}(G)\le 5$. Moreover, the bound is sharp.
\end{corollary}
\begin{proof}
     Let $G$ be a cactus graph. Note that a cactus graph $G$ is a block-cactus graph such that $\omega(G) \le 3$. Thus, from \Cref{thm:block_cactus_graph}, we have $2 \le \chi_{o}(G)\le 5$.

\end{proof}
Next, we characterize cactus graphs based on their odd chromatic number. Let \(G\) be a connected cactus graph, and let \(\mathcal{C}(G)\) denote the family of blocks that are cycles in \(G\). Furthermore, let $F$ be the spanning subgraph of $G$ whose edges are precisely the bridges of $G$. Since \(G\) is a cactus graph, \(F\) is a forest.



\begin{theorem}\label{thm:cactus-three}
Let $G$ be a connected cactus graph. Then $\chi_o(G)\leq 3$ if and only if for every cycle block $C\in\mathcal{C}(G)$ there exists a set $A_C\subseteq E(C)$ such that:
\begin{enumerate}
\item[\textup{(i)}] For every $C\in\mathcal{C}(G)$, $ |E(C)|-2|A_C|\equiv0\pmod 3.$

\item[\textup{(ii)}] For every vertex $v$ incident with no bridge,
 the number of cycle blocks $C$ containing $v$ for which the two
 edges of $C$ incident with $v$ are either both in $A_C$ or both
 outside $A_C$ is odd.
\end{enumerate}
\end{theorem}

\begin{proof}
Suppose $\varphi:V(G)\to\mathbb Z_3$ is an odd $3$-coloring.
For each cycle block $C=v_1v_2\cdots v_nv_1$, traverse its vertices
in this order and put
\[
 A_C=\bigl\{v_iv_{i+1}\in E(C):
   \varphi(v_{i+1})-\varphi(v_i)\equiv-1\pmod3\bigr\},
\]
where indices are read modulo $n$.
Since $\varphi$ is proper, every color difference around this cycle
is either $1$ or $-1$ modulo $3$. Summing around the cycle gives
\[
 0\equiv \sum_{i=1}^n
       \bigl(\varphi(v_{i+1})-\varphi(v_i)\bigr)
   \equiv (n-|A_C|)-|A_C|
   =n-2|A_C|\pmod3.
\]
Thus (i) holds.

Let $v \in V(G)$ be incident with no bridge.  Because $G$ is a cactus graph,
every edge at $v$ belongs to exactly one cycle block, and each such
block contributes two incident edges.  In particular, $d_G(v)$ is even.
For a cycle block $C$ containing $v$, its two neighbors of $v$ have
different colors precisely when the two incident edges of $C$ are
both in $A_C$ or both outside $A_C$.  Such a cycle contributes one
neighbor of each of the two colors other than $\varphi(v)$.
Otherwise, the two neighbors have the same color and the cycle
contributes either zero or two neighbors of either color.
Consequently, modulo $2$, the number of neighbors in either one of
the two other color classes equals the number of cycle blocks
specified in (ii).  As $d_G(v)$ is even and $\varphi$ is odd, both
neighbor-color classes have odd size.  Therefore the number in (ii)
is odd, as required.

\medskip
Conversely, suppose the edge sets $A_C$ satisfy (i) and (ii).  For each vertex $v \in V(G)$,
write $t(v)$ for the number of cycle blocks containing $v$ in which
the two edges incident with $v$ are both selected or both unselected.

We first assign each bridge to one of its endpoints.  Let $m(v)$
denote the number of bridges assigned to $v$.  We require
\begin{equation}\label{eq:bridge-parity_1}
 t(v)+m(v) \equiv 1\pmod2
 \quad\text{whenever $d_G(v)$ is even and $d_F(v)>0$.}
\end{equation}
Such an assignment always exists.  Indeed,
\[d_G(v)=d_F(v)+2\,|\{C\in\mathcal C(G):v\in V(C)\}|,\]
so $d_G(v)$ and $d_F(v)$ have the same parity.
For each non-trivial tree component of the bridge forest $F$,
choose a leaf as root.  Process the other vertices from the leaves
toward the root.  At each non-root vertex of even degree in $F$,
assign its parent bridge to the appropriate endpoint so that
\eqref{eq:bridge-parity_1} holds; at a vertex of odd degree in $F$,
choose arbitrarily.  The root has degree $1$ in $F$, so it requires
no parity condition.  If $v$ is isolated in $F$, then (ii) already
implies $t(v)$ is odd.

Next, prescribe a difference $\delta(x,y)\in\{1,-1\}\subseteq\mathbb Z_3$
for every ordered pair of adjacent vertices, with
$\delta(y,x)=-\delta(x,y)$.
For the chosen cyclic traversal of
$C=v_1\cdots v_nv_1$, set
\[
 \delta(v_i,v_{i+1})= 
 \begin{cases}
   -1,&v_iv_{i+1}\in A_C,\\
    1,&v_iv_{i+1}\notin A_C.
 \end{cases}
\]
If a bridge $xy$ was assigned to $x$, set $\delta(x,y)=1$.
By (i), the sum of the prescribed differences around every cycle
block is zero modulo $3$. As every cycle of a cactus graph is a
cycle block, the difference prescriptions are consistent along all
paths.  Choose a root vertex $r$, put $\varphi(r)=0$, and define
$\varphi(v)$ by summing $\delta$ along any path from $r$ to $v$.
Thus $\varphi:V(G)\to\mathbb Z_3$ is well-defined, and it is proper
because every edge has difference $1$ or $-1$.

For any vertex $v$, let $ p(v)=\bigl|\{u\in N_G(v):
       \varphi(u)-\varphi(v)\equiv1\pmod3\}\bigr|.$
Each cycle block counted by $t(v)$ contributes exactly one neighbor
to $p(v)$; each other incident cycle block contributes zero or two.
Each bridge assigned to $v$ contributes exactly one.  Therefore
\begin{equation}\label{eq:color-parity_2}
 p(v)\equiv t(v)+m(v)\pmod2.
\end{equation}
If $d_G(v)$ is even and $d_F(v)>0$, then \eqref{eq:bridge-parity_1}
and \eqref{eq:color-parity_2} show that $p(v)$ is odd.  If
$d_G(v)$ is even and $d_F(v)=0$, condition (ii) gives the same
conclusion.  In either case, $d_G(v)-p(v)$ is also odd, so $v$
satisfies the odd condition.
If $d_G(v)$ is odd, at least one of $p(v)$ and $d_G(v)-p(v)$ is odd
automatically.  Hence every vertex satisfies the odd condition,
and $\varphi$ is an odd $3$-coloring of $G$.  Thus $\chi_o(G)\leq3$.

\end{proof}

From Observation~\ref{obs:oddchromaticnumber_2}, if \(G\) is a connected graph with at least two vertices, then \(\chi_o(G)=2\) if and only if \(G\) is bipartite and every vertex of \(G\) has an odd degree. Due to this fact and Theorem~\ref{thm:cactus-three}, we have the following corollary:



\begin{corollary}\label{cor:three}
Let $G$ be a connected cactus graph with at least two vertices.
Then $\chi_o(G)=3$ if and only if conditions \textup{(i)} and
\textup{(ii)} of Theorem~\ref{thm:cactus-three} hold and $G$ is not
a bipartite graph in which every vertex has odd degree.
\end{corollary}

\begin{proof}
By Theorem~\ref{thm:cactus-three}, we have $\chi_o(G)\leq 3$
if and only if conditions \textup{(i)} and \textup{(ii)} hold.
Moreover, by Observation~\ref{obs:oddchromaticnumber_2},
the additional condition excludes $\chi_o(G)=2$.
Hence, $\chi_o(G)=3$.
\end{proof}


\begin{corollary}\label{cor:four}
Let $G$ be a connected cactus graph with at least two vertices.
Then $\chi_o(G)=4$ if and only if
\begin{enumerate}
\item $G$ has a block not isomorphic to $C_5$; and
\item There is no family $(A_C:C\in\mathcal C(G))$ satisfying
conditions \textup{(i)} and \textup{(ii)} of
Theorem~\ref{thm:cactus-three}.
\end{enumerate}
\end{corollary}

\begin{proof}
Every cactus graph is outerplanar.  Kashima and Zhu proved that
a connected outerplanar graph is odd $4$-colorable if and only if
it has a block not isomorphic to $C_5$~ \cite{kashima2024odd}.
Thus the first condition is equivalent to $\chi_o(G)\leq4$.
By Theorem~\ref{thm:cactus-three}, the second condition is
equivalent to $\chi_o(G)>3$.  Together these conditions give
$\chi_o(G)=4$, and the converse follows from the same equivalences.
\end{proof}

\medskip
\noindent Next, we give examples of cactus graphs having odd chromatic number $3$ and $4$.

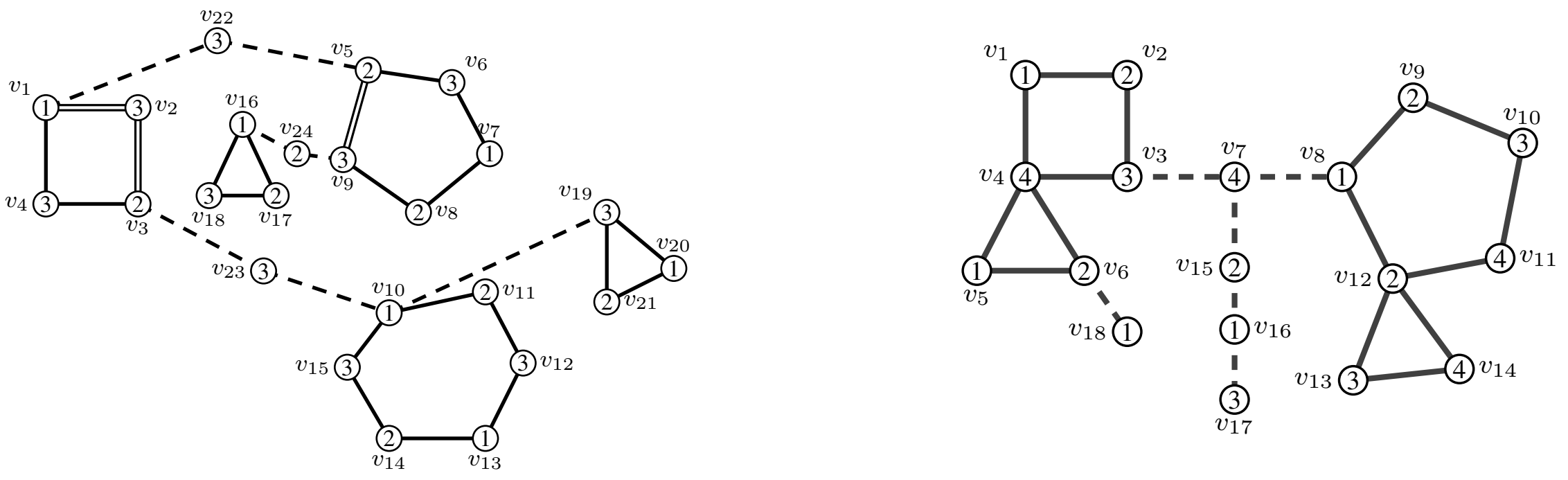
\begin{figure}[htbp]
\centering

\begin{minipage}{0.48\textwidth}
\centering

\begin{tikzpicture}[
    scale=0.80,
    transform shape,
    x=0.010cm, y=-0.010cm,
    every node/.style={font=\scriptsize},
    vertex/.style={
        circle, draw=black, fill=white,
        line width=0.50pt,
        minimum size=3mm,
        inner sep=0pt,
        text=black,
        font=\scriptsize
    },
    every label/.style={
        font=\scriptsize,
        text=black,
        inner sep=1pt
    },
    cycle/.style={
        draw=black,
        line width=1.0pt
    },
    bridge/.style={
        draw=black,
        dashed,
        dash pattern=on 4pt off 3pt,
        line width=1.0pt
    },
    marked/.style={
        draw=black,
        double,
        double distance=0.8pt,
        line width=0.6pt
    }
]

\coordinate (p1) at (130,120);
\coordinate (p2) at (240,120);
\coordinate (p3) at (240,235);
\coordinate (p4) at (130,235);
\coordinate (p5) at (515,75);
\coordinate (p6) at (615,90);
\coordinate (p7) at (660,175);
\coordinate (p8) at (575,245);
\coordinate (p9) at (485,182);
\coordinate (p10) at (540,365);
\coordinate (p11) at (655,340);
\coordinate (p12) at (700,425);
\coordinate (p13) at (655,515);
\coordinate (p14) at (540,515);
\coordinate (p15) at (490,430);
\coordinate (p16) at (365,140);
\coordinate (p17) at (405,225);
\coordinate (p18) at (325,225);
\coordinate (p19) at (800,245);
\coordinate (p20) at (880,312);
\coordinate (p21) at (800,352);
\coordinate (p22) at (335,40);
\coordinate (p23) at (390,315);
\coordinate (p24) at (430,175);

\draw[marked] (p1) -- (p2);
\draw[marked] (p2) -- (p3);
\draw[cycle] (p3) -- (p4);
\draw[cycle] (p4) -- (p1);

\draw[cycle] (p5) -- (p6);
\draw[cycle] (p6) -- (p7);
\draw[cycle] (p7) -- (p8);
\draw[cycle] (p8) -- (p9);
\draw[marked] (p9) -- (p5);

\draw[cycle] (p10) -- (p11);
\draw[cycle] (p11) -- (p12);
\draw[cycle] (p12) -- (p13);
\draw[cycle] (p13) -- (p14);
\draw[cycle] (p14) -- (p15);
\draw[cycle] (p15) -- (p10);

\draw[cycle] (p16) -- (p17);
\draw[cycle] (p17) -- (p18);
\draw[cycle] (p18) -- (p16);

\draw[cycle] (p19) -- (p20);
\draw[cycle] (p20) -- (p21);
\draw[cycle] (p21) -- (p19);

\draw[bridge] (p1) -- (p22);
\draw[bridge] (p22) -- (p5);
\draw[bridge] (p3) -- (p23);
\draw[bridge] (p23) -- (p10);
\draw[bridge] (p9) -- (p24);
\draw[bridge] (p24) -- (p16);
\draw[bridge] (p10) -- (p19);


\node[vertex,label=above left:$v_1$] at (p1) {1};
\node[vertex,label=above:$v_7$] at (p7) {1};
\node[vertex,label=above:$v_{10}$] at (p10) {1};
\node[vertex,label=below:$v_{13}$] at (p13) {1};
\node[vertex,label=above:$v_{16}$] at (p16) {1};
\node[vertex,label=above:$v_{20}$] at (p20) {1};

\node[vertex,label=below:$v_3$] at (p3) {2};
\node[vertex,label=above left:$v_5$] at (p5) {2};
\node[vertex,label=right:$v_8$] at (p8) {2};
\node[vertex,label=right:$v_{11}$] at (p11) {2};
\node[vertex,label=below:$v_{14}$] at (p14) {2};
\node[vertex,label=below:$v_{17}$] at (p17) {2};
\node[vertex,label=right:$v_{21}$] at (p21) {2};
\node[vertex,label=above:$v_{24}$] at (p24) {2};

\node[vertex,label=right:$v_2$] at (p2) {3};
\node[vertex,label=left:$v_4$] at (p4) {3};
\node[vertex,label=above right:$v_6$] at (p6) {3};
\node[vertex,label=below:$v_9$] at (p9) {3};
\node[vertex,label=right:$v_{12}$] at (p12) {3};
\node[vertex,label=left:$v_{15}$] at (p15) {3};
\node[vertex,label=below:$v_{18}$] at (p18) {3};
\node[vertex,label=above left:$v_{19}$] at (p19) {3};
\node[vertex,label=above:$v_{22}$] at (p22) {3};
\node[vertex,label=left:$v_{23}$] at (p23) {3};

\end{tikzpicture}

\medskip
{\small (a) An odd $3$-color cactus graph.}

\end{minipage}
\hfill
\begin{minipage}{0.43\textwidth}
\centering

\begin{tikzpicture}[
    scale=0.90,
    transform shape,
    x=0.012cm, y=-0.012cm,
    every node/.style={font=\scriptsize},
    vertex/.style={
        circle, draw=black, fill=white,
        line width=0.65pt,
        minimum size=3mm,
        inner sep=0pt,
        text=black,
        font=\scriptsize
    },
    every label/.style={
        font=\scriptsize,
        text=black,
        inner sep=1pt
    },
    cycle/.style={
        draw=black!75,
        line width=1.55pt
    },
    bridge/.style={
        draw=black!75,
        dashed,
        dash pattern=on 4pt off 3pt,
        line width=1.55pt
    }
]

\coordinate (p1) at (65,45);
\coordinate (p2) at (155,45);
\coordinate (p3) at (155,135);
\coordinate (p4) at (65,135);
\coordinate (p5) at (22,218);
\coordinate (p6) at (117,218);
\coordinate (p7) at (250,135);
\coordinate (p8) at (345,135);
\coordinate (p9) at (408,65);
\coordinate (p10) at (505,105);
\coordinate (p11) at (485,207);
\coordinate (p12) at (390,225);
\coordinate (p13) at (355,315);
\coordinate (p14) at (449,305);
\coordinate (p15) at (250,215);
\coordinate (p16) at (250,270);
\coordinate (p17) at (250,332);
\coordinate (p18) at (155,272);

\draw[cycle] (p1) -- (p2);
\draw[cycle] (p2) -- (p3);
\draw[cycle] (p3) -- (p4);
\draw[cycle] (p4) -- (p1);

\draw[cycle] (p4) -- (p5);
\draw[cycle] (p5) -- (p6);
\draw[cycle] (p6) -- (p4);

\draw[cycle] (p8) -- (p9);
\draw[cycle] (p9) -- (p10);
\draw[cycle] (p10) -- (p11);
\draw[cycle] (p11) -- (p12);
\draw[cycle] (p12) -- (p8);

\draw[cycle] (p12) -- (p13);
\draw[cycle] (p13) -- (p14);
\draw[cycle] (p14) -- (p12);

\draw[bridge] (p3) -- (p7);
\draw[bridge] (p7) -- (p8);
\draw[bridge] (p7) -- (p15);
\draw[bridge] (p15) -- (p16);
\draw[bridge] (p16) -- (p17);
\draw[bridge] (p6) -- (p18);


\node[vertex,label=above left:$v_1$] at (p1) {1};
\node[vertex,label=below:$v_5$] at (p5) {1};
\node[vertex,label=above left:$v_8$] at (p8) {1};
\node[vertex,label=right:$v_{16}$] at (p16) {1};
\node[vertex,label=left:$v_{18}$] at (p18) {1};

\node[vertex,label=above right:$v_2$] at (p2) {2};
\node[vertex,label=right:$v_6$] at (p6) {2};
\node[vertex,label=above:$v_9$] at (p9) {2};
\node[vertex,label=left:$v_{12}$] at (p12) {2};
\node[vertex,label=left:$v_{15}$] at (p15) {2};

\node[vertex,label=above right:$v_3$] at (p3) {3};
\node[vertex,label=above:$v_{10}$] at (p10) {3};
\node[vertex,label=left:$v_{13}$] at (p13) {3};
\node[vertex,label=below:$v_{17}$] at (p17) {3};

\node[vertex,label=left:$v_4$] at (p4) {4};
\node[vertex,label=above:$v_7$] at (p7) {4};
\node[vertex,label=right:$v_{11}$] at (p11) {4};
\node[vertex,label=right:$v_{14}$] at (p14) {4};

\end{tikzpicture}

\medskip
{\small (b) An odd $4$-color cactus graph.}

\end{minipage}

\caption{Two connected cactus graphs. The numbers inside
the vertices represent their assigned colors, and dashed
edges represent bridges. In \textup{(a)}, double edges
indicate the selected sets
$A_{C_1}=\{(1,2),(2,3)\}$ and
$A_{C_2}=\{(5,9)\}$, while
$A_{C_3}=A_{C_4}=A_{C_5}=\varnothing$.
In \textup{(b)}, no edge set $A_C$ for the $4$-cycle
satisfies both conditions \textup{(i)} and \textup{(ii)}.}

\label{fig:cactus_odd_34_coloring}

\end{figure}

\begin{observation}
\label{obs:oddchromaticnumber_5}
Let $G$ be a connected cactus graph. Then $\chi_o(G)=5$ if and only if every block of $G$ is isomorphic to $C_5$.
\end{observation}

\begin{proof}
Since every cactus graph is outerplanar, the result of Petruševski and Škrekovski \cite{CARO2022392} implies that $\chi_o(G)\leq 5$. Moreover, Kashima and Zhu \cite{kashima2024odd} proved that a connected outerplanar graph admits
an odd $4$-coloring if and only if it contains a block that is not isomorphic to $C_5$. Consequently, $\chi_o(G)=5$ if and only if every block of $G$ is isomorphic to $C_5$.
\end{proof}


We summarize all the results proved above through the following corollary:
\begin{corollary}
Let $G$ be a connected cactus graph with at least two
vertices. Then the odd chromatic number of $G$ is given by
\[
\chi_o(G)=
\begin{cases}
2, & \begin{aligned}[t]
     &\text{if $G$ is bipartite and every vertex has odd degree};\\
     \end{aligned}\\[3mm]
3, & \begin{aligned}[t]
     &\text{if conditions \textup{(i)} and \textup{(ii)}
       of Theorem~\ref{thm:cactus-three} hold and $G$ is not}\\
     &\text{bipartite with all vertices of odd degree};\\
     \end{aligned}\\[3mm]
4, & \begin{aligned}[t]
     &\text{if $G$ has a block not isomorphic to $C_5$ and no family}\\
     &\text{$(A_C:C\in\mathcal C(G))$ satisfies conditions \textup{(i)} and
       \textup{(ii)} of Theorem~\ref{thm:cactus-three};}
     \end{aligned}\\[3mm]
5, & \begin{aligned}[t]
     &\text{if every block of $G$ is isomorphic to $C_5$.}
     \end{aligned}
\end{cases}
\]
\end{corollary}


\section{Block Graphs}
For a graph $G$, a maximal induced subgraph of $G$ without a cut vertex is called a block of $G$. We use $B, B_1, B_2, \ldots$ to denote blocks of a graph $G$. By the definition of blocks, we note that if $B_1$ and $B_2$ are two blocks, then $|V(B_1) \cap V(B_2)| \le 1$.  
For our purpose of establishing a bound, we use the block-cut tree of a block graph $G$. 

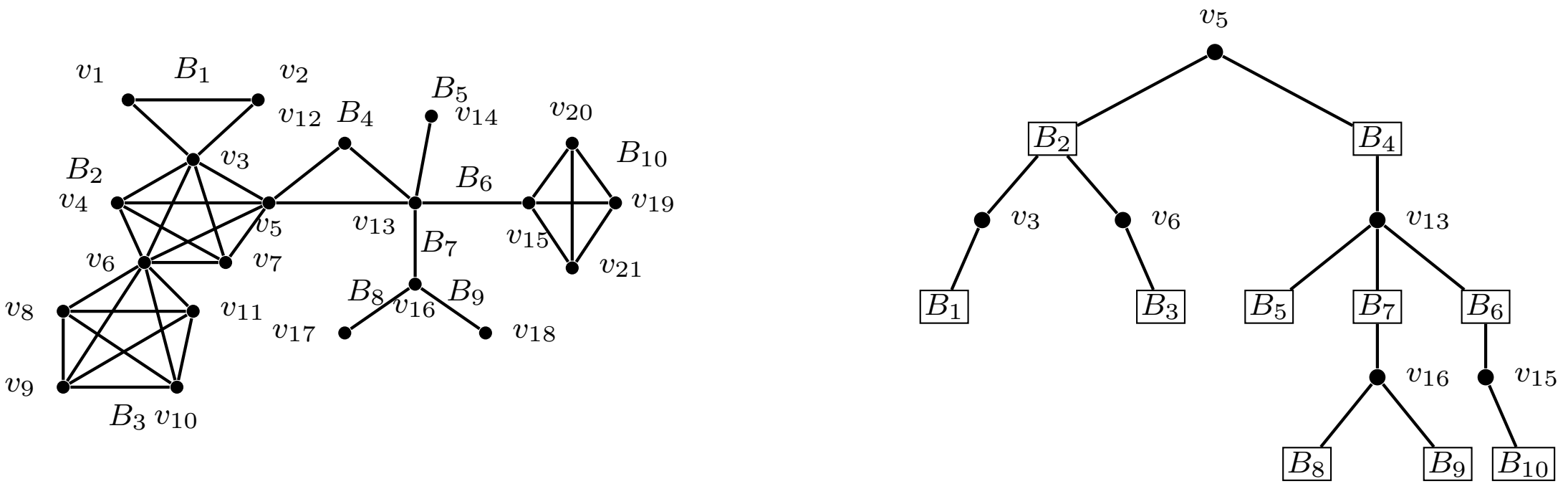
\begin{figure}[ht!]
\centering

\begin{minipage}[c]{0.48\textwidth}
\centering

\begin{tikzpicture}[
    scale=0.50,
    vertex/.style={
        circle,
        fill=black,
        minimum size=1.2mm,
        inner sep=0pt
    },
    label/.style={
        font=\scriptsize,
        inner sep=1pt
    },
    edge/.style={
        draw=black,
        thick
    },
    blocklabel/.style={
        draw=none,
        font=\scriptsize,
        inner sep=1pt
    }
]


\node[vertex] (v1) at (0.8,5.8) {};
\node[vertex] (v2) at (3.2,5.8) {};
\node[vertex] (v3) at (2.0,4.7) {};

\node[vertex] (v4) at (0.6,3.9) {};
\node[vertex] (v5) at (3.4,3.9) {};
\node[vertex] (v6) at (1.1,2.8) {};
\node[vertex] (v7) at (2.6,2.8) {};

\node[vertex] (v8)  at (-0.4,1.9) {};
\node[vertex] (v9)  at (-0.4,0.5) {};
\node[vertex] (v10) at (1.7,0.5) {};
\node[vertex] (v11) at (2.0,1.9) {};

\node[vertex] (v12) at (4.8,5.0) {};
\node[vertex] (v13) at (6.1,3.9) {};
\node[vertex] (v14) at (6.4,5.5) {};
\node[vertex] (v15) at (8.2,3.9) {};

\node[vertex] (v16) at (6.1,2.4) {};
\node[vertex] (v17) at (4.8,1.5) {};
\node[vertex] (v18) at (7.4,1.5) {};

\node[vertex] (v19) at (9.8,3.9) {};
\node[vertex] (v20) at (9,5.0) {};
\node[vertex] (v21) at (9,2.7) {};


\node[label, above left=2pt and 3pt of v1] {$v_1$};
\node[label, above right=2pt and 3pt of v2] {$v_2$};
\node[label, right=4pt of v3] {$v_3$};

\node[label, left=4pt of v4] {$v_4$};
\node[label, below=1pt of v5] {$v_5$};
\node[label, left=4pt of v6] {$v_6$};
\node[label, right=4pt of v7] {$v_7$};

\node[label, left=4pt of v8] {$v_8$};
\node[label, left=4pt of v9] {$v_9$};
\node[label, below=3pt of v10] {$v_{10}$};
\node[label, right=4pt of v11] {$v_{11}$};

\node[label, above left=2pt and 3pt of v12] {$v_{12}$};
\node[label, below left=1pt and 2pt of v13] {$v_{13}$};
\node[label, right=3pt of v14] {$v_{14}$};
\node[label, below=4pt of v15] {$v_{15}$};

\node[label, below=1pt of v16] {$v_{16}$};
\node[label, left=4pt of v17] {$v_{17}$};
\node[label, right=4pt of v18] {$v_{18}$};

\node[label, right=1pt of v19] {$v_{19}$};
\node[label, above=3pt of v20] {$v_{20}$};
\node[label, right=4pt of v21] {$v_{21}$};


\draw[edge] (v1) -- (v2);
\draw[edge] (v1) -- (v3);
\draw[edge] (v2) -- (v3);

\node[blocklabel] at (2.0,6.35) {$B_1$};


\draw[edge] (v3) -- (v4);
\draw[edge] (v3) -- (v5);
\draw[edge] (v3) -- (v6);
\draw[edge] (v3) -- (v7);

\draw[edge] (v4) -- (v5);
\draw[edge] (v4) -- (v6);
\draw[edge] (v4) -- (v7);

\draw[edge] (v5) -- (v6);
\draw[edge] (v5) -- (v7);
\draw[edge] (v6) -- (v7);

\node[blocklabel] at (0,4.5) {$B_2$};


\draw[edge] (v8) -- (v9);
\draw[edge] (v8) -- (v10);
\draw[edge] (v8) -- (v11);
\draw[edge] (v9) -- (v10);
\draw[edge] (v9) -- (v11);
\draw[edge] (v10) -- (v11);

\draw[edge] (v6) -- (v8);
\draw[edge] (v6) -- (v9);
\draw[edge] (v6) -- (v10);
\draw[edge] (v6) -- (v11);

\node[blocklabel] at (0.8,-0.05) {$B_3$};


\draw[edge] (v5) -- (v12);
\draw[edge] (v5) -- (v13);
\draw[edge] (v12) -- (v13);

\node[blocklabel] at (5,5.55) {$B_4$};


\draw[edge] (v13) -- (v14);
\draw[edge] (v13) -- (v15);
\draw[edge] (v13) -- (v16);

\node[blocklabel] at (6.75,6.0) {$B_5$};
\node[blocklabel] at (7.2,4.35) {$B_6$};
\node[blocklabel] at (6.55,3.15) {$B_7$};


\draw[edge] (v16) -- (v17);
\draw[edge] (v16) -- (v18);

\node[blocklabel] at (5.2,2.25) {$B_8$};
\node[blocklabel] at (7.05,2.25) {$B_9$};


\draw[edge] (v15) -- (v19);
\draw[edge] (v15) -- (v20);
\draw[edge] (v15) -- (v21);

\draw[edge] (v19) -- (v20);
\draw[edge] (v19) -- (v21);
\draw[edge] (v20) -- (v21);

\node[blocklabel] at (10.3,4.8) {$B_{10}$};

\end{tikzpicture}

\end{minipage}
\hfill
\begin{minipage}[c]{0.48\textwidth}
\centering

\begin{tikzpicture}[
    scale=0.50,
    every node/.style={
        font=\scriptsize
    },
    block/.style={
        rectangle,
        draw=black,
        minimum width=4mm,
        minimum height=3mm,
        inner sep=1pt
    },
    cut/.style={
        circle,
        fill=black,
        minimum size=1.5mm,
        inner sep=0pt
    },
    label/.style={
        font=\scriptsize,
        inner sep=1pt
    },
    edge/.style={
        draw=black,
        thick
    },
    weight/.style={
        draw=none,
        font=\scriptsize\bfseries,
        inner sep=1pt
    }
]


\node[cut] (v5) at (0,5.8) {};

\node[label, above=3pt of v5] {$v_5$};


\node[block] (B2) at (-3,4.2) {$B_2$};
\node[block] (B4) at (3,4.2) {$B_4$};

\draw[edge] (v5) -- (B2);
\draw[edge] (v5) -- (B4);


\node[cut] (v3) at (-4.3,2.7) {};
\node[cut] (v6) at (-1.7,2.7) {};

\node[label, right=4pt of v3] {$v_3$};
\node[label, right=4pt of v6] {$v_6$};

\draw[edge] (B2) -- (v3);
\draw[edge] (B2) -- (v6);

\node[block] (B1) at (-5.0,1.1) {$B_1$};
\node[block] (B3) at (-1.0,1.1) {$B_3$};

\draw[edge] (v3) -- (B1);
\draw[edge] (v6) -- (B3);


\node[cut] (v13) at (3,2.7) {};

\node[label, right=4pt of v13] {$v_{13}$};

\draw[edge] (B4) -- (v13);


\node[block] (B5) at (1.0,1.1) {$B_5$};
\node[block] (B7) at (3.0,1.1) {$B_7$};
\node[block] (B6) at (5.0,1.1) {$B_6$};

\draw[edge] (v13) -- (B5);
\draw[edge] (v13) -- (B7);
\draw[edge] (v13) -- (B6);


\node[cut] (v16) at (3.0,-0.2) {};

\node[label, right=4pt of v16] {$v_{16}$};

\draw[edge] (B7) -- (v16);


\node[block] (B8) at (1.7,-1.8) {$B_8$};
\node[block] (B9) at (4.3,-1.8) {$B_9$};

\draw[edge] (v16) -- (B8);
\draw[edge] (v16) -- (B9);


\node[cut] (v15) at (5.0,-0.2) {};

\node[label, right=4pt of v15] {$v_{15}$};

\draw[edge] (B6) -- (v15);


\node[block] (B10) at (5.7,-1.8) {$B_{10}$};

\draw[edge] (v15) -- (B10);

\end{tikzpicture}

\end{minipage}

\caption{A block graph $G$ and its corresponding rooted block-cut tree $T_G$.}
\label{fig:block-graph-block-cut-tree}

\end{figure}

Let $G$ be a block graph, $\mathcal{B}(G)$ denote the set of blocks of $G$, and  $X$ denote the set of cut vertices of $G$. The \emph{block-cut tree} of $G$, denoted by $T_G$, is a tree with vertex set $X\cup\mathcal{B}(G)$, in which a cut vertex $x\in X$ is adjacent to a block $B\in\mathcal{B}(G)$ if and only if $x\in V(B)$. Note that each leaf vertex of the tree $T_G$ denotes an end block of $G$. 
We consider the block tree $ T_ {G}$ rooted at a vertex $r$ such that $r$ corresponds to a cut vertex of $G$. In our proofs, we consider the vertices of $T_{G}$ in a BFS-ordering starting with vertex $r$. An illustration of a block graph and its corresponding block tree is given in Fig.~\ref{fig:block-graph-block-cut-tree}. A BFS-ordering of block tree $T_G$ given in Fig.~\ref{fig:block-graph-block-cut-tree} is given by $\{v_5, B_2, B_4, v_3,v_6,v_{13}, B_1, B_3, B_5, B_7, B_6, v_{16},v_{15}, B_8, B_9, B_{10}\}$.

\begin{theorem}
\label{thm:block_graph}
    Let $G$ be a block graph. Then $\omega(G) \le \chi_{o}(G)\le \omega(G)+1$.
\end{theorem}
 
\begin{proof}
    The proof is similar to the proof of \Cref{thm:block_cactus_graph}. For the sake of completeness, we illustrate the proof again. Let $G$ be a block graph. The lower bound is obvious from the fact that every block graph $H$ is a perfect graph, $\chi(H) = \omega(H)$ and $\chi_o(H) \ge \chi(H)$. To establish the upper bound, we use induction on the number of blocks in the block graph $G$.


    Suppose $G$ has only one block. Thus, $G$ is a complete graph on $n$ vertices, then $\chi_o(G)= n =\omega(G)$. This establishes the base case. Next, we assume that the result holds for any block graph having at most $r-1$ blocks for $r\ge 2$.



    Now consider a block graph $G$ having $r$ blocks. Recall that any block graph $G$ has at least two end blocks. Let $B$ be an end block with cut vertex $v$. Notably, $B$ is a clique. Let $S=(V(G)-V(B))\cup \{v\}$ and $H= G[S]$. Note that the graph $H$ is a block graph having $r-1$ blocks. Therefore, by the inductive hypothesis, $H$ admits an odd coloring $f$ using at most $\omega(H)+1$ colors. Since $\omega(H) \le \omega(G)$, $H$ admits an odd coloring using at most $\omega(G)+1$ colors. As $f$ is an odd coloring of $H$, there exists a color $i \in [\omega(G)+1]$ such that $|f^{-1}(i) \cap N(v)|$ is odd. Also, assume that $f(v) = j$. Next, we extend the odd coloring $f$ of $H$ to an odd coloring $f'$ of $G$ such that $f'$ uses at most $\omega(G)+1$ colors.  
    
    We note that $|V(B)| \le \omega(G)$ and no vertex of $V(B) \setminus \{v\}$ has a neighbor in $V(G) \setminus V(B)$. Thus, we extend the coloring $f$ of $H$ to an odd coloring $f'$ of $G$ as follows: $f'(u)=f(u)$ for each vertex $u \in V(G) \setminus (V(B) \setminus \{v\})$. Assign all vertices in $V(B) \setminus \{v\} $ a distinct color from the set $[\omega(G)+1] \setminus \{i,j\}$. Now, we claim that the coloring $f'$ is an odd coloring of $G$. Noting the fact that no vertex of $V(B) \setminus \{v\}$ has a neighbor in $V(G) \setminus V(B)$ and $f$ is a odd coloring of $H$, for any pair of vertices $u,w \in V(G) \setminus V(B)$, if $uw \in E(G)$, $f(u) \neq f(w)$. Also, for each vertex $u \in V(G) \setminus V(B)$, there exists a color $k \in [\omega(G)+1]$ such that $|f'^{-1}(k) \cap N_G(u)|$ is odd. 
    
    Next, note that no vertex in $V(B) \setminus \{v\}$ is assigned color $i$ or $j$. Thus $|f'^{-1}(i) \cap N_G(v)|$ remains odd. Also, for each vertex $u \in V(B) \setminus \{v\}$, $|f'^{-1}(j) \cap N_G(u)| = 1$. Precisely, $v$ is the only vertex colored $j$ in the neighborhood of each vertex $u \in V(B) \setminus \{v\}$. This concludes that $f'$ is an odd coloring of $G$ that uses at most $\omega(G)+1$ colors.


 
\end{proof}

\vt{From the above theorem, we have that, for any block graph $G$, $\chi_{o}(G) \in \{\omega(G), \omega(G)+1\}$. Moreover the proposed bounds are tight. For example a complete graph on $5$ vertices $K_5$ achieves the lower bound while two $K_5$ sharing exactly one vertex achieve the upper bound. Next, we characterize block graphs $G$ depending on the odd chromatic number of $G$. Consider a block graph $G$. Let $B$ and $v$ be a block and a cut vertex in $G$, respectively. We say that {\em $B$ is incident on $v$} if $v\in V(B)$.}

\begin{theorem}\label{thm:block_graph_charectarization}
    \vt{Let $G$ be a block graph. Then $\chi_o(G)=\omega(G)+1$ if and only if there exists a cut vertex $v$ such that the following is satisfied:
    \begin{enumerate}
        \item the size of each block incident to $v$ is $\omega(G)$, and 
        \item the number of blocks incident to $v$ is even.
    \end{enumerate}}   
\end{theorem}

\begin{proof}
   Let $G$ be a block graph. Due to~\Cref{thm:block_graph}, $\chi_{o}(G) \in \{\omega(G) , \omega(G)+1\}$. First, we assume that $\chi_{o}(G) = \omega(G)+1$. We need to claim that there exists a cut vertex $v$ such that the following is satisfied:
    \begin{enumerate}
        \item the size of each block incident to $v$ is $\omega(G)$, and 
        \item the number of blocks incident to $v$ is even.
    \end{enumerate}
    In order to prove this, we claim that if there does not exist any cut vertex with both properties together, then $\chi_{o}(G)=\omega(G)$. To achieve this we describe a coloring procedure that uses at most $\omega(G)$ colors.
Thus, for every cut vertex $v$, either some block incident with $v$ has
order less than $\omega$, or the number of blocks incident with $v$ is
odd. We prove that $G$ has an odd coloring using $\omega$ colors.

We now construct a rooted block-cut tree $T_G$ rooted at a cut vertex $r$. We consider the vertices of $T_G$ in the BFS order and assign the color to the vertices of $V(G)$. Intuitively, if we are assigning colors to a cut vertex $v$ in our proposed procedure, then at the same time we decide the color of the vertices of each block $B$ adjacent to $v$. The coloring procedure is as follows: start by assigning an arbitrary color to the cut vertex $v_r$ corresponding to the root vertex~$r$. Suppose $f(v_r) = i$. Let $t_1,t_2, \ldots, t_k$ be the block vertices adjacent to $r$ and $B_{t_1},B_{t_2}, \ldots, B_{t_k}$ be the corresponding blocks. 
\begin{itemize}
    \item {\bf If $|B_{t_j}| = \omega(G)$ for each $j \in [k]$}.
    
    In this case, according to the assumption, $k$ is odd. Hence, for each $j \in [k]$, every uncolored vertex of block $B_j$ is assigned a distinct color from $[\omega(G)] \setminus \{i\}$. That is, for each $j \in [k]$, assign color to uncolored vertices in $V(B_j)$ from the set $[\omega(G)] \setminus \{i\}$ such that $f(w) \neq f(u)$ for all edge $wu\in E(B_j)$. Note that when $f$ restricted to blocks $B_{t_1},B_{t_2}, \ldots, B_{t_k}$ is an odd coloring.

    \item {\bf If there exists a $j \in [k]$ such that $|B_{t_j}| < \omega(G)$}.
    
    In this case, we first assign the color to the uncolored vertices of the block other than $B_j$. For each $j' \in [k] \setminus \{j\}$, assign color to uncolored vertices in $V(B_{j'})$ from the set $[\omega(G)] \setminus \{i\}$ such that $f(w) \neq f(u)$. If there exists a color $i' \in [\omega(G)]$ such that $|f^{-1}(i') \cap N(v_r)|$ is odd, then assign color to uncolored vertices in $V(B_j)$ from the set $[\omega(G)] \setminus \{i,i'\}$ such that $f(w) \neq f(u)$. This is indeed possible as the number of uncolored vertices in $B_j$ are at most $\omega(G)-2$. Otherwise, whichever color from $[\omega(G)]\setminus\{i\}$ is assign to uncolored vertices in \(V(B_j)\) is odd. Consequently, $f$ when restricted to blocks $B_{t_1},B_{t_2}, \ldots, B_{t_k}$ is an odd coloring.
\end{itemize}

We note that, in this procedure, the color of a cut vertex is decided while coloring the vertices of the corresponding block. Let $t$ be any cut vertex. Let $t_1$ be the parent of $t$. Note that $t_1$ is a block vertex corresponding to some block, say $B_{t_1}$. \vt{The color of $v_{t}$ is already defined while coloring the vertices of $B_{t_1}$. Suppose, $f(v_{t}) = i$.} Let $t_2,t_3, \ldots, t_k$ be the block vertices adjacent to $t$ other than $t_1$ and $B_{t_2},B_{t_3}, \ldots, B_{t_k}$ be the corresponding blocks.

\begin{itemize}
    \item {\bf If $|B_{t_j}| = \omega(G)$ for each $j \in [k]$}.
    
    In this case, according to the assumption, $k$ is odd. Hence, for each $j \in [k]$, every uncolored vertex of block $B_j$ is assigned a distinct color from $[\omega(G)] \setminus \{i\}$. That is, for each $j \in [k]$, assign color to uncolored vertices in $V(B_j)$ from the set $[\omega(G)] \setminus \{i\}$ such that $f(w) \neq f(u)$. Hence any color from the set $[\omega(G)] \setminus \{i\}$ occurs an odd number of times in $N_G(v_t)$. Thus $f$ is an odd coloring.
    
    \item {\bf If there exists a $j \in [k]$ such that $|B_{t_j}| < \omega(G)$}.
    
    We find the maximum index $j$ such that $|B_{t_j}| < \omega(G)$. Suppose $j \ge 2$. That is, there exist a child block having size less than $\omega(G)$. In this case, we first assign the color to the uncolored vertices of the block other than $B_j$. For each $j' \in [k] \setminus \{j\}$, assign color to uncolored vertices in $V(B_{j'})$ from the set $[\omega(G)] \setminus \{i\}$ such that $f(w) \neq f(u)$. If there exists a color $i' \in [\omega(G)]$ such that $|f^{-1}(i') \cap N(v_t)|$ is odd, then assign color to uncolored vertices in $V(B_j)$ from the set $[\omega(G)] \setminus \{i,i'\}$ such that $f(w) \neq f(u)$. This is indeed possible as the number of uncolored vertices in $B_j$ are at most $\omega(G)-2$. Otherwise, whichever color from $[\omega(G)]\setminus\{i\}$ is assign to uncolored vertices in \(V(B_j)\) is odd.

    If $j=1$, then all child blocks have size $\omega(G)$. In this case, for each $j' \in [k]$, assign color to uncolored vertices in $V(B_{j'})$ from the set $[\omega(G)] \setminus \{i\}$ such that $f(w) \neq f(u)$. We claim that there exists a color $i' \in [\omega(G)]$ such that $|f^{-1}(i') \cap N(v_t)|$ is odd. We claim this by considering the following cases:
    \begin{itemize}
        \item Suppose $k$ is odd. Hence, the number of child blocks incident to vertex $v_t$ is even. Hence, any color assigned to a vertex $V(B_{t_1}) \setminus \{v_t\}$ is appearing odd number of times in $N_{G}(v_t)$.
        \item If $k$ is even. Then, the number of child blocks incident to vertex $v_t$ is odd. Hence, any color in the set $[\omega(G)] \setminus f(V(B_{t_1})$ is appearing odd number of times in $N_{G}(v_t)$, since $|V(B_{t_1})| < \omega(G)$. Thus $f$ is an odd coloring.
    \end{itemize}
    
\end{itemize}

    Conversely, suppose there exists a cut vertex $v$ such that 
    \begin{enumerate}
        \item the size of each block incident to $v$ is $\omega(G)$, and 
        \item the number of blocks incident to $v$ is even.
    \end{enumerate}

    On the contrary, suppose that there exists an odd coloring $f$ of $G$ that uses at most $\omega(G)$ colors. Let $f(v)=i$ and $B_1, B_2, \ldots, B_{r}$ be the block incident to the vertex $v$. Due to our assumption, we note that $r$ is even and $|V(B_k)|$ is $\omega(G)$ for each $k \in [r]$. Since every odd coloring is a proper coloring, for any two distinct vertices $u$ and $w$, $f(u) \neq f(w)$ in any block $B_k$, $k \in [r]$. Thus $f$ uses $\omega(G)-1$ colors other than color $i$ to color the vertices of $V(B_k) \setminus \{v\}$ for each $k \in [r]$. This concludes that all $\omega(G)-1$ colors other than $i$ appear even number of times in $N(v)$. This contradicts the fact that $f$ is an odd coloring of $G$. Therefore, the theorem follows.



\end{proof}


\section{Proper Interval graph}

A graph $G$ is a \emph{proper interval graph} if it admits an interval representation $\mathcal{I}=\{I_v:v\in V(G)\}$ on the real line such that, for every pair of distinct vertices $u,v\in V(G),~uv\in E(G)\Longleftrightarrow I_u\cap I_v\neq\emptyset$, and no interval in $\mathcal{I}$ properly contains another. Proper interval graphs are characterized by a vertex ordering called {\em bi-compatible ordering} (BCO). Let $G = (V,E)$ be a graph. An ordering $\alpha = \{v_1,v_2, \ldots, v_n\}$ is called bi-compatible ordering if for $i<j$, $v_iv_j \in E(G)$ implies $v_kv_i,v_kv_j \in E(G)$ for each $i<k<j$. It is known that a graph $G$ is a proper interval graph if and only if it admits a BCO~\cite{jamison1982elimination}.  Define
$N^-(v_i)=N(v_i)\cap\{v_1,\ldots,v_{i-1}\}$ and $N^+(v_i)=N(v_i) \cap\{v_{i+1}, \ldots,v_n\}$.

We make the following important observation about the maximum degree of a proper interval graph.

\begin{observation}
     Every vertex $v$ of a proper interval graph $G$ satisfies $\deg_G(v)\leq 2\omega(G)-2$.
\end{observation}

\begin{proof}
Let $\alpha=(v_1,v_2,\ldots,v_n)$ be a BCO of $G$, and let $v_i\in V(G)$. By the BCO property, $G[N^-(v_i)\cup\{v_i\}]$ and $G[N^+(v_i)\cup\{v_i\}]$
are cliques. Hence, $|N^-(v_i)|+1\leq\omega(G)$ and $|N^+(v_i)|+1\leq\omega(G)$.
Therefore $\deg_G(v_i)=|N^-(v_i)|+|N^+(v_i)|\leq(\omega(G)-1)+(\omega(G)-1) =2\omega(G)-2$. Since $v_i$ is arbitrary, $\Delta(G)\leq2\omega(G)-2$.
\end{proof}

Next, we establish bounds on the odd chromatic number for proper interval graphs, followed by characterization depending on the odd chromatic number.
\begin{theorem}
\label{thm:proper_interval_graph}
Let $G$ be a proper interval graph. Then, $\omega(G)\le \chi_o (G)\le \omega(G)+1$.
\end{theorem}

\begin{proof}
Let $G=(V,E)$ be a proper interval graph with bi-compatible ordering (BCO) $\alpha=(v_1,v_2,\ldots,v_n)$. It is known that the vertices appearing in a BCO of a proper interval graph form a Hamiltonian path. The lower bound is immediate. Since every clique of $G$ must receive distinct colors in any proper coloring, we have $\chi_o(G)\ge \omega(G)$. It remains to establish the upper bound. To establish this, define a coloring $f:V(G)\rightarrow[\omega+1]$ by $f(v_i)=1+((i-1) \bmod(\omega+1))$. Thus, the colors are assigned cyclically as $\{1,2,\ldots,\omega, \omega+1\}$. 

We first prove that $f$ is a proper coloring. Suppose, for contradiction, that $v_iv_j\in E(G)$ for some $i<j$ and $f(v_i)=f(v_j)$. By the definition of $f$, $j-i\equiv0\pmod{(\omega+1)}$, and hence $j-i\geq\omega+1$. Therefore, $|\{v_i,v_{i+1},\ldots,v_j\}|=j-i+1\geq\omega+2$. Since $\alpha$ is a BCO and $v_iv_j\in E(G)$, the vertices $\{v_i,v_{i+1},\ldots,v_j\}$ induce a clique. Consequently, $\omega(G)\geq j-i+1\geq\omega(G)+2$, which is a contradiction. Hence, $f$ is a proper coloring.

It remains to prove the odd coloring condition. For every $2\leq i\leq n-1$, the preceding neighbors of $v_i$ with $v_i$ form a clique. Therefore, $|N(v_i)\cap\{v_1,\ldots,v_{i-1}\}|\leq\omega-1$. Similarly, $|N(v_i)\cap\{v_{i+1},\ldots,v_n\}|\leq\omega-1$. By the definition of $f$, $f(v_{i-1})\equiv f(v_i)-1\pmod{(\omega+1)}$ and $f(v_{i+1})\equiv f(v_i)+1\pmod{(\omega+1)}$. Since each side of $v_i$ contains at most $\omega-1$ neighbors and the coloring is periodic with period $\omega+1$, the color $f(v_{i-1})$ occurs exactly once in $N(v_i)$, namely at $v_{i-1}$. Similarly, the color $f(v_{i+1})$ occurs exactly once in $N(v_i)$, namely at $v_{i+1}$. Hence, $|N(v_i)\cap f^{-1}(f(v_{i-1}))|=1$ and $|N(v_i)\cap f^{-1}(f(v_{i+1}))|=1$. Thus, at least two colors occur an odd number of times in $N(v_i)$. For $v_1$, all its neighbors occur after $v_1$, and $|N(v_1)|\leq\omega-1$. Hence, $|N(v_1)\cap f^{-1}(f(v_2))|=1$. Similarly, for $v_n$, all its neighbors occur before $v_n$, and $|N(v_n)|\leq\omega-1$. Hence, $|N(v_n)\cap f^{-1}(f(v_{n-1}))|=1$.

Therefore, for every $v\in V(G)$, there exists a color $c\in[\omega+1]$ such that $|N(v)\cap f^{-1}(c)|\equiv1\pmod 2$. Thus, $f$ is an odd coloring of $G$, and consequently, $\chi_o(G)\leq\omega(G)+1$. Combining this with the lower bound $\chi_o(G)\ge \omega(G)$, we conclude that $\omega(G)\le \chi_o(G)\le \omega(G)+1$.
\end{proof}

\begin{figure}[ht]
    \centering
    \begin{tikzpicture}[
        vertex/.style={
            circle,
            draw=black,
            fill=white,
            minimum size=1mm,
            inner sep=0.5pt,
            font=\scriptsize\bfseries
        },
        edge/.style={thick}
    ]

    \node[vertex] (v1)  at (0,0)    {1};
    \node[vertex] (v2)  at (0.8,0)  {2};
    \node[vertex] (v3)  at (1.6,0)  {3};
    \node[vertex] (v4)  at (2.4,0)  {4};
    \node[vertex] (v5)  at (3.2,0)  {5};
    \node[vertex] (v6)  at (4.0,0)  {1};
    \node[vertex] (v7)  at (4.8,0)  {2};
    \node[vertex] (v8)  at (5.6,0)  {3};
    \node[vertex] (v9)  at (6.4,0)  {4};
    \node[vertex] (v10) at (7.2,0)  {5};

    \draw[edge] (v1) to[bend left=0] (v2);
    \draw[edge] (v1) to[bend left=35] (v3);

    \draw[edge] (v2) to[bend left=0] (v3);
    \draw[edge] (v2) to[bend left=35] (v4);

    \draw[edge] (v3) to[bend left=0] (v4);
    \draw[edge] (v3) to[bend left=28] (v5);
    \draw[edge] (v3) to[bend left=45] (v6);

    \draw[edge] (v4) to[bend left=0] (v5);
    \draw[edge] (v4) to[bend left=28] (v6);

    \draw[edge] (v5) to[bend left=0] (v6);
    \draw[edge] (v5) to[bend left=35] (v7);

    \draw[edge] (v6) to[bend left=0] (v7);
    \draw[edge] (v6) to[bend left=35] (v8);
    \draw[edge] (v6) to[bend left=45] (v9);

    \draw[edge] (v7) to[bend left=0] (v8);
    \draw[edge] (v7) to[bend left=30] (v9);

    \draw[edge] (v8) to[bend left=0] (v9);
    \draw[edge] (v8) to[bend left=28] (v10);

    \draw[edge] (v9) to[bend left=0] (v10);

    \node[below=2pt] at (v1)  { $v_1$};
    \node[below=2pt] at (v2)  { $v_2$};
    \node[below=2pt] at (v3)  { $v_3$};
    \node[below=2pt] at (v4)  { $v_4$};
    \node[below=2pt] at (v5)  { $v_5$};
    \node[below=2pt] at (v6)  { $v_6$};
    \node[below=2pt] at (v7)  { $v_7$};
    \node[below=2pt] at (v8)  { $v_8$};
    \node[below=2pt] at (v9)  { $v_9$};
    \node[below=2pt] at (v10) { $v_{10}$};

    \end{tikzpicture}

    \caption{Example of an interval graph attaining the upper bound in \Cref{thm:proper_interval_graph}. Here, $\chi_o(G)=5$, while $\omega(G)=4$ and $\Delta(G)=2\omega(G)-2=6$.}
    \label{fig:cyclic-six-coloring}
\end{figure}
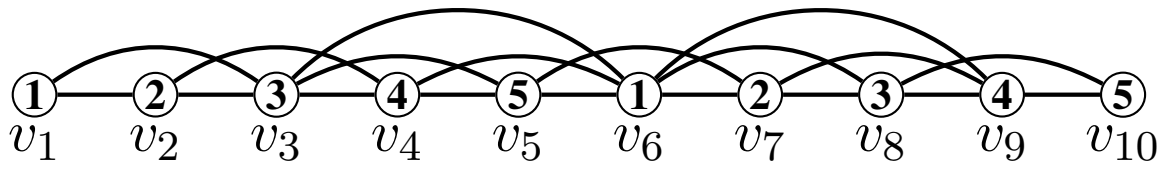

From \Cref{thm:proper_interval_graph}, we note that the odd chromatic number of a proper interval graph is sandwiched between $\omega(G)$ and $\omega(G)+1$. Here it is important to remark that both the bounds are tight. In Fig.~\ref{fig:cyclic-five-coloring} and Fig.~\ref{fig:cyclic-six-coloring}, we illustrate proper interval graphs attaining lower and upper bounds, respectively.

\begin{figure}[ht]
    \centering
    \begin{tikzpicture}[
        vertex/.style={
            circle,
            draw=black,
            fill=white,
            minimum size=1mm,
            inner sep=0.5pt,
            font=\scriptsize\bfseries
        },
        edge/.style={thick}
    ]

    \node[vertex] (v1)  at (0,0)    {1};
    \node[vertex] (v2)  at (0.8,0)  {2};
    \node[vertex] (v3)  at (1.6,0)  {3};
    \node[vertex] (v4)  at (2.4,0)  {4};
    \node[vertex] (v5)  at (3.2,0)  {1};
    \node[vertex] (v6)  at (4.0,0)  {2};
    \node[vertex] (v7)  at (4.8,0)  {3};
    \node[vertex] (v8)  at (5.6,0)  {4};
    \node[vertex] (v9)  at (6.4,0)  {1};
    \node[vertex] (v10) at (7.2,0)  {2};

    \draw[edge] (v1) to[bend left=0] (v2);
    \draw[edge] (v1) to[bend left=35] (v3);

    \draw[edge] (v2) to[bend left=0] (v3);
    \draw[edge] (v2) to[bend left=35] (v4);

    \draw[edge] (v3) to[bend left=0] (v4);
    \draw[edge] (v3) to[bend left=28] (v5);
    \draw[edge] (v3) to[bend left=45] (v6);

    \draw[edge] (v4) to[bend left=0] (v5);
    \draw[edge] (v4) to[bend left=28] (v6);

    \draw[edge] (v5) to[bend left=0] (v6);
    \draw[edge] (v5) to[bend left=35] (v7);

    \draw[edge] (v6) to[bend left=0] (v7);
    \draw[edge] (v6) to[bend left=35] (v8);

    \draw[edge] (v7) to[bend left=0] (v8);
    \draw[edge] (v7) to[bend left=30] (v9);
    \draw[edge] (v7) to[bend left=45] (v10);

    \draw[edge] (v8) to[bend left=0] (v9);
    \draw[edge] (v8) to[bend left=28] (v10);

    \draw[edge] (v9) to[bend left=0] (v10);

    \node[below=2pt] at (v1)  { $v_1$};
    \node[below=2pt] at (v2)  { $v_2$};
    \node[below=2pt] at (v3)  { $v_3$};
    \node[below=2pt] at (v4)  { $v_4$};
    \node[below=2pt] at (v5)  { $v_5$};
    \node[below=2pt] at (v6)  { $v_6$};
    \node[below=2pt] at (v7)  { $v_7$};
    \node[below=2pt] at (v8)  { $v_8$};
    \node[below=2pt] at (v9)  { $v_9$};
    \node[below=2pt] at (v10) { $v_{10}$};

    \end{tikzpicture}

    \caption{Example of an interval graph attaining the lower bound in \Cref{thm:proper_interval_graph}. Here, $\chi_o(G)=4$, while $\omega(G)=4$ and $\Delta(G)=5<6=2\omega(G)-2$.}
    \label{fig:cyclic-five-coloring}
\end{figure}
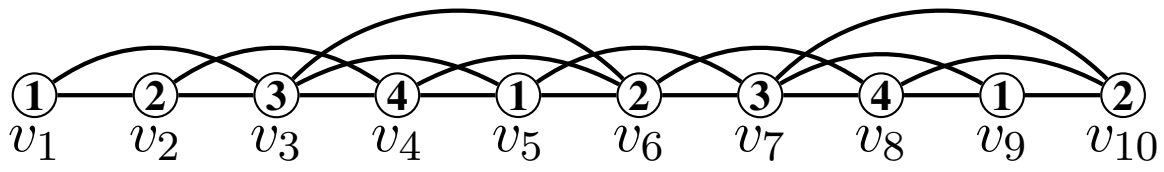

    Our next theorem characterizes the class of proper interval graphs depending on the odd chromatic number. 
    
\begin{theorem}\label{thm_proper_char}
Let $G$ be a proper interval graph. Then $\chi_o(G)=\omega(G)+1$ if and only if there exists a vertex $v\in V(G)$ such that $\deg_G(v)=2\omega(G)-2$. Otherwise, $\chi_o(G)=\omega(G)$.
\end{theorem}
\begin{proof}
Let $G=(V,E)$ be a proper interval graph with bi-compatible ordering (BCO) $\alpha=(v_1,v_2,\ldots,v_n)$. Let $\omega=\omega(G)$. By the previous \Cref{thm:proper_interval_graph}, we have $\omega\leq\chi_o(G) \leq \omega+1$. Suppose $\chi_o(G)=\omega+1$. We claim that there exists a vertex $v\in V(G)$ such that $\deg_G(v)=2\omega-2$.  To prove this, define a coloring $f:V(G) \rightarrow [\omega]$ by $f(v_i)=1+((i-1)\bmod\omega)$. Since $\alpha$ is a BCO, for every edge $v_iv_j\in E(G)$ with $i<j$, the vertices $v_i,v_{i+1},\ldots,v_j$ induce a clique. Hence $j-i+1\leq\omega$, and thus $j-i<\omega$, Since the coloring is periodic with period $\omega$, implying $f(v_i)\neq f(v_j)$. Therefore, $f$ is a proper $\omega$-coloring. 




For any $v_i\in V(G)$, let $p_i=|N^-(v_i)|$ and $q_i=|N^+(v_i)|$, where $N^-(v_i)=N(v_i)\cap\{v_1,\ldots,v_{i-1}\}$ and $N^+(v_i)=N(v_i)\cap\{v_{i+1},\ldots,v_n\}$. By the BCO property, $G[N^-(v_i)\cup{v_i}]$ and $G[N^+(v_i)\cup{v_i}]$ are cliques, then $p_i,q_i\leq\omega-1$, and hence $\deg_G(v_i) \leq2\omega-2$. Moreover, under $f$, the colors on $N^-(v_i)$ and $N^+(v_i)$ are respectively consecutive colors preceding and succeeding $f(v_i)$, that is $f(v_i)-p_i, \ldots, f(v_i)-1$ and $f(v_i)+1, \ldots,f(v_i)+q_i$ modulo $\omega$ . Thus, every color occurs at most twice in $N(v_i)$. If $\deg_G(v_i)\leq2\omega-3$, then there exists some color $j \in [\omega]$ such that $|N(v_i) \cap f^{-1}(j)| \equiv1 \pmod 2$; otherwise the set of colors appearing on $N^-(v_i)$ is equal to the set of colors appearing on $N^+(v_i)$. Hence, $\{f(v_i)-p_i, \ldots,f(v_i)-1\}=\{f(v_i)+1, \ldots,f(v_i)+q_i\}$ modulo $\omega$. This equality is possible only when $p_i=q_i=\omega-1$, and hence $\deg_G(v_i)=2\omega-2$, a contradiction. Therefore, $\deg_G(v_i) \leq2 \omega-3$ implies that $v_i$ satisfies the odd coloring condition.

Suppose $\deg_G(v)\leq2\omega-3$ for every $v\in V(G)$, then $f$ is an odd $\omega$-coloring, Together with $\chi_o(G)\geq\omega$, we obtain $\chi_o(G)=\omega$, a contradiction. Hence, there exists $v\in V(G)$ such that $\deg_G(v)=2\omega-2$.

Conversely, suppose that $\deg_G(v)=2\omega-2$ for some $v\in V(G)$. Then $p_v=q_v= \omega-1$. By the BCO property, $G[N^-(v)\cup\{v\}]$ and $G[N^+(v)\cup\{v\}]$ are cliques of order $\omega$.  For any proper $\omega$-coloring $g$, each of these cliques receives all $\omega$ colors. Hence, $g(N^-(v))=g(N^+(v))=[\omega]\setminus{g(v)}$. Therefore, $|N(v)\cap g^{-1}(j)|=2$ for every $j\in [\omega]\setminus{g(v)}$. Thus, every color occurs an even number of times in $N(v)$, so $v$ violates the odd coloring condition. Hence, $\chi_o(G)>\omega$, and therefore $\chi_o(G)=\omega+1$. Thus, $\chi_o(G)=\omega+1$ if and only if there exists $v\in V(G)$ such that $\deg_G(v)=2\omega-2$; otherwise, $\chi_o(G)=\omega$.

\end{proof}

\subsection{A linear time algorithm for odd coloring in proper interval graphs} 
In this section, we present a linear-time algorithm that, given a proper interval graph $G$ and a BCO $\alpha=(v_1,v_2,\ldots,v_n)$ of $G$, computes an odd coloring of $G$ with minimum colors. We remark that clique size of a proper interval graph can be computed in linear-time~\cite{clique_prop_int}. The algorithm is as follows:

\begin{algorithm}[H]
\caption{Odd Coloring of a Proper Interval Graph}
\label{alg:odd-coloring-proper-interval}
\KwIn{A proper interval graph $G=(V,E)$ and BCO $\alpha=(v_1,v_2,\ldots,v_n)$ of $G$}
\KwOut{An odd coloring $f$ of $G$}

Initialize: $\chi_{o}(G) = \emptyset$ \;

\If{there exists a vertex such that ${\rm deg}_{G}(v) = 2 \omega(G)-2$}{
     $\chi_{o}(G) = \omega(G)+1$
}
\Else{
    $\chi_{o}(G) = \omega(G)$
}


\If{$\chi_o(G)=\omega$}{
    \For{$i\leftarrow 1$ \KwTo $n$}{
        $f(v_i)\leftarrow 1+((i-1)\bmod\omega)$\;
    }
}
\Else{
    \For{$i\leftarrow 1$ \KwTo $n$}{
        $f(v_i)\leftarrow 1+((i-1)\bmod(\omega+1))$\;
    }
}
\Return{$f$}\;
\end{algorithm}

\medskip
The correctness of \Cref{alg:odd-coloring-proper-interval} follows from the \Cref{thm:proper_interval_graph} and \Cref{thm_proper_char}. Next, we analyze the time complexity of the proposed algorithm. We note that the maximum clique size $\omega(G)$ of a proper interval graph can be computed in linear time. Moreover, checking whether there exists a vertex of degree with exactly $2\omega(G)-2$ can be done in linear time. Also, our coloring procedure is linear-time. This concludes the linearity of the proposed algorithm.


\section{Bipartite Circle Graphs}

A connected bipartite distance-hereditary graph is a connected bipartite graph $G$ in which the distance between any two vertices remains unchanged in every connected induced subgraph containing them. Equivalently, such a graph can be constructed from a single vertex by repeatedly adding pendant vertices or false twins. A bipartite circle graph is a bipartite graph that can be represented as the intersection graph of chords of a circle, where two vertices are adjacent if and only if their corresponding chords intersect. Every distance-hereditary graph is a circle graph; consequently, the class of bipartite distance-hereditary graphs is a proper subclass of the class of bipartite circle graphs. Davis et al.\,in \cite{Davies2025OddColoring} study the problem in circle and related graph classes. They pose the following question as an open problem 

\medskip
\noindent \textbf{Problem~\cite{Davies2025OddColoring}} {\em  Does every bipartite circle graph have odd
chromatic number at most $4$?}

\medskip
Below, we give a construction of a bipartite circle graph $G$ such that $\chi_{o}(G) = 6$, thereby disproving the above statement.

\begin{proposition}\label{prop:counterexample}
There is a connected bipartite circle graph $G$ on ten vertices with $\chi_o(G)=5.$ 
\end{proposition}

\begin{proof}
Let $G$ be the bipartite graph with bi-partition $X=\{a,a',b,c,d\},\text{ and } Y=\{p,p',q,q',r\}$. Fig.~\ref{fig:graph} displays $G$ with a proper odd $5$-coloring.

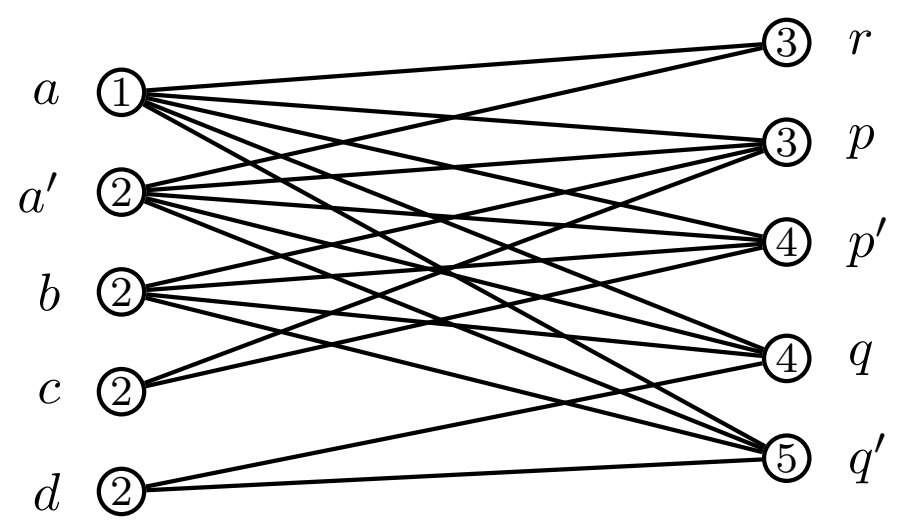
\begin{figure}[htbp]
\centering
\begin{tikzpicture}[
  scale=0.55,
  vertex/.style={
    circle,
    draw=black,
    fill=white,
    thick,
    minimum size=3mm,
    inner sep=0pt,
    font=\scriptsize
  },
  edge/.style={draw=black,line width=0.8pt},
  lab/.style={font=\small}
]

\node[vertex] (a)  at (-4,  2.4) {$1$};
\node[vertex] (ap) at (-4,  1.2) {$2$};
\node[vertex] (b)  at (-4,  0.0) {$2$};
\node[vertex] (c)  at (-4, -1.2) {$2$};
\node[vertex] (d)  at (-4, -2.4) {$2$};

\node[lab,left=3pt] at (a.west)  {$a$};
\node[lab,left=3pt] at (ap.west) {$a'$};
\node[lab,left=3pt] at (b.west)  {$b$};
\node[lab,left=3pt] at (c.west)  {$c$};
\node[lab,left=3pt] at (d.west)  {$d$};

\node[vertex] (r)  at (4,  3.0) {$3$};
\node[vertex] (p)  at (4,  1.8) {$3$};
\node[vertex] (pp) at (4,  0.6) {$4$};
\node[vertex] (q)  at (4, -0.8) {$4$};
\node[vertex] (qp) at (4, -2.0) {$5$};

\node[lab,right=3pt] at (r.east)  {$r$};
\node[lab,right=3pt] at (p.east)  {$p$};
\node[lab,right=3pt] at (pp.east) {$p'$};
\node[lab,right=3pt] at (q.east)  {$q$};
\node[lab,right=3pt] at (qp.east) {$q'$};

\foreach \v in {r,p,pp,q,qp} {
  \draw[edge] (a) -- (\v);
  \draw[edge] (ap) -- (\v);
}

\foreach \v in {p,pp,q,qp} {
  \draw[edge] (b) -- (\v);
}

\foreach \v in {p,pp} {
  \draw[edge] (c) -- (\v);
}

\foreach \v in {q,qp} {
  \draw[edge] (d) -- (\v);
}

\end{tikzpicture}
\caption{A bipartite circle graph $G$ with
$\chi_o(G)=5$. The numbers inside the vertices
represent color of respective vertex.}
\label{fig:graph}
\end{figure}

To verify that $G$ is a circle graph, place the following double-occurrence
word around a circle, in cyclic order:
\[
r,\ a',\ a,\ r,\ b,\ c,\ p',\ p,\ c,\ d,\ q',\ q,\ d,\ b,\ a,\ a',\ q,\ q',\ p,\ p'.
\]
Join the two occurrences of each symbol by a chord. Two chords cross
if and only if their endpoints alternate around the circle, a direct
check gives exactly the adjacency sets specified above. Thus $G$ is a
bipartite circle graph.

We next prove that $\chi_o(G)\geq5$. Suppose, for a contradiction,
that $\varphi$ is a proper odd coloring of $G$ using at most four colors.
The three vertices $r,c,d$ have degree two, with $N(r)=\{a,a'\}, N(c)=\{p,p'\}, d N(d)=\{q,q'\}.$ At a degree-two vertex, the odd-neighborhood condition requires its
two neighbors to receive different colors. Therefore,
\[
\varphi(a)\neq\varphi(a'),\qquad
\varphi(p)\neq\varphi(p'),\qquad
\varphi(q)\neq\varphi(q').
\]
Every vertex of $\{p,p',q,q'\}$ is adjacent to both $a$ and $a'$.
Properness consequently forbids these four vertices from using either
of the two distinct colors $\varphi(a),\varphi(a')$.
At most two colors remain, say $\alpha$ and $\beta$. Since the colors
on $p,p'$ differ, as do those on $q,q'$, we have $\{\varphi(p),\varphi(p')\}=\{\alpha,\beta\}
=\{\varphi(q),\varphi(q')\}.$ But $N(b)=\{p,p',q,q'\}$, so each of $\alpha$ and $\beta$ occurs
exactly twice in $N(b)$, and no other color occurs there. This
contradicts the odd-neighborhood condition at $b$. Hence
$\chi_o(G)\geq5$.

For the reverse inequality, define $\psi:V(G)\to\{1,2,3,4,5\}$ by
\begin{align*}
\psi(a)&=1,\\
\psi(a')=\psi(b)=\psi(c)=\psi(d)&=2,\\
\psi(p)=\psi(r)&=3,\\
\psi(p')=\psi(q)&=4,\\
\psi(q')&=5.
\end{align*}
This coloring is proper. The colors in the neighborhoods of $a$ and
$a'$ are $3,3,4,4,5$; at $b$ they are $3,4,4,5$; and at each of
$p,p',q,q'$ they are $1,2,2,2$. Finally, the two neighbors of each
of $c,d,r$ have different colors. Thus every vertex has a color
occurring an odd number of times in its neighborhood, proving
$\chi_o(G)\leq5$.
Combining the inequalities yields $\chi_o(G)=5$.
\end{proof}

\begin{remark}
    In fact, the graph above is bipartite distance-hereditary. It may be obtained from the cycle $a-p-b-q-a$ by adjoining pendant vertices $r,c,d$ at $a,p,q$, respectively, and then false twins $a',p',q'$ of $a,p,q$, respectively. 
\end{remark}

\begin{theorem}\label{thm:bcr6}
Every bipartite circle graph $G$ satisfies $\chi_o(G)\leq 6.$
\end{theorem}

\begin{proof}
We first establish the upper bound. 
Since the odd chromatic number of a disjoint
union is the maximum of the odd chromatic
numbers of its components, it suffices to
consider a connected bipartite circle graph
$G$ containing at least one edge.

By the characterization of De. Fraysseix
\cite{de1981local}, For every connected bipartite circle graph there exists a
connected plane graph $H$ with a spanning
tree $T$ such that $G\cong F(H,T),$ where $F(H,T)$ is the fundamental graph
of $H$ with respect to $T$.

We identify the bi-partition of $G$ with $X=E(T), Y=E(H)\setminus E(T).$ For $e\in X$ and $f\in Y$, the vertices
$e$ and $f$ are adjacent in $G$ if and
only if $e$ belongs to the fundamental
cycle of $f$ with respect to $T$. Since $G$ is connected and has an edge,
it has no isolated vertices. Therefore,
$H$ has no loops or bridges: a loop
would correspond to an isolated vertex
in $Y$, whereas a bridge would correspond
to an isolated vertex in $X$. Now we construct a proper odd coloring of
$G$ using six colors.

\medskip
\noindent\textbf{Coloring the vertices in $X$.}

By the Four Color Theorem, there exists
a proper vertex coloring $g:V(H)\longrightarrow\mathbb F_2^2,$ where $\mathbb F_2^2=\{00,01,10,11\}.$ For each tree edge $e=uv\in E(T)$,
define
\[
\lambda(e)=g(u)\oplus g(v),
\]
where $\oplus$ denotes componentwise
addition modulo $2$. Since $g$ is a proper coloring, $g(u)\neq g(v),$ and hence $\lambda(e)\in\{01,10,11\}$ (if $g(u)= g(v),$ then $g(u)\oplus g(v)=00$). Assign colors $1,2,3$ to the vertices
of $X$ according to these three nonzero
vectors.

We claim that every vertex of $Y$
satisfies the odd coloring condition. Let $f=uv\in Y$. Since $f$ is not a
tree edge, the fundamental cycle of
$f$ consists of $f$ together with the
unique $u$--$v$ path $P_T(u,v)$ in $T$. Thus, $N_G(f)=E(P_T(u,v)).$ Consequently,
\[
\begin{aligned}
\bigoplus_{e\in N_G(f)}\lambda(e)
&=
\bigoplus_{xy\in E(P_T(u,v))}
\bigl(g(x)\oplus g(y)\bigr)\\
&=g(u)\oplus g(v).
\end{aligned}
\]
The second equality follows because
every internal vertex of the path
appears exactly twice in the XOR sum. Since $f=uv$ is an edge of $H$ and
$g$ is proper, $g(u)\oplus g(v)\neq 00.$ Therefore, $\bigoplus_{e\in N_G(f)}\lambda(e)
\neq 00.$

If each of the three colors occurred
an even number of times in $N_G(f)$,
then every vector in $\{01,10,11\}$
would occur an even number of times,
forcing the XOR sum to be $00$. This is a contradiction.

Hence at least one color among
$\{1,2,3\}$ occurs an odd number
of times in $N_G(f)$. Thus every vertex of $Y$ satisfies
the odd coloring condition.

\medskip
\noindent\textbf{Coloring the vertices in $Y$.}

Let $H^*$ denote the planar dual
of $H$. Since $T$ is a spanning tree of $H$,
the dual edges corresponding to
$E(H)\setminus E(T)$ form a spanning
tree $T^*$ of $H^*$. Since $H$ is bridgeless, its planar
dual $H^*$ has no loops. By the Four Color Theorem, there
exists a proper vertex coloring $h:V(H^*)\longrightarrow\mathbb F_2^2.$ Equivalently, $h$ is a proper coloring
of the faces of $H$ using four colors. For each non-tree edge $f\in Y$,
let $F_1$ and $F_2$ be the two faces
of $H$ incident with $f$. Define
\[
\mu(f)=h(F_1)\oplus h(F_2).
\]
Since $H$ has no bridges, $F_1$
and $F_2$ are distinct adjacent
faces. Therefore, $\mu(f)\in\{01,10,11\}.$ Assign colors $4,5,6$ to the vertices
of $Y$ according to these three
non-zero vectors.

We claim that every vertex of $X$
satisfies the odd coloring condition. Let $e\in X$, and let $F_1,F_2$ be
the two faces incident with $e$. The neighbors of $e$ in $G$ are
exactly the non-tree edges belonging
to the fundamental cut of $T$
determined by deleting $e$. By planar duality, the duals of
these non-tree edges form the unique
$F_1$--$F_2$ path in $T^*$. Consequently, the XOR sum along this
dual path telescopes, giving
\[
\bigoplus_{f\in N_G(e)}\mu(f)
=
h(F_1)\oplus h(F_2).
\]
Since $H$ has no bridges, the faces
$F_1$ and $F_2$ are distinct and
adjacent in $H^*$.
As $h$ is proper, $h(F_1)\oplus h(F_2)\neq 00.$ Therefore,
\[
\bigoplus_{f\in N_G(e)}\mu(f)
\neq 00.
\]

As before, if each of the colors
$4,5,6$ occurred an even number
of times in $N_G(e)$, the XOR sum
would be $00$, a contradiction. Hence at least one color among
$\{4,5,6\}$ occurs an odd number
of times in $N_G(e)$. Thus every vertex of $X$ also
satisfies the odd coloring condition.

\medskip
\noindent\textbf{Properness of the coloring.}

The vertices of $X$ receive colors
only from $\{1,2,3\}$, whereas the
vertices of $Y$ receive colors only
from $\{4,5,6\}$. Since $G$ is bipartite with
bi-partition $(X,Y)$, every edge
has one endpoint in $X$ and the
other in $Y$. Therefore, the coloring is proper.

Since every non-isolated vertex
satisfies the odd neighborhood
condition, we obtain a proper odd
coloring using at most six colors. Consequently, $\chi_o(G)\leq 6.$

\end{proof}

\begin{corollary}
\label{cor:cbdhg6}
    Every connected bipartite distance-hereditary graph $G$ satisfies $\chi_o(G)\leq 6.$ Moreover, the bound is sharp.
\end{corollary}

\begin{proof}
Since every connected bipartite distance-hereditary graph
is a bipartite circle graph, the inequality
$\chi_o(G)\leq 6$ follows directly from
\Cref{thm:bcr6}. The sharpness of this bound
is established in \Cref{thm:sharpnessbound_of_BCG}.
\end{proof}


\subsection{Sharpness of the Bound}

\begin{theorem}
\label{thm:sharpnessbound_of_BCG}
There exists a connected bipartite distance-hereditary
graph $G$ such that $\chi_o(G)=6$. Consequently,
there exists a bipartite circle graph with odd
chromatic number $6$.
\end{theorem}

\begin{proof}
\begin{figure}[htbp]
\centering
\begin{tikzpicture}[
    scale=0.55,
    vertex/.style={
        circle,
        draw=black,
        fill=white,
        thick,
        minimum size=3mm,
        inner sep=0pt,
        font=\scriptsize
    },
    edge/.style={
        draw=black!75,
        line width=0.65pt
    },
    lab/.style={
        font=\scriptsize,
        fill=white,
        inner sep=1pt
    }
]


\coordinate (t) at (-7,0);

\coordinate (d)  at (-4, 3.5);
\coordinate (a1) at (-4, 2.1);
\coordinate (a2) at (-4, 0.8);
\coordinate (b1) at (-4,-0.8);
\coordinate (b2) at (-4,-2.1);
\coordinate (e)  at (-4,-3.5);

\coordinate (r)  at (3.5, 3.5);
\coordinate (p1) at (3.5, 2.1);
\coordinate (p2) at (3.5, 0.8);
\coordinate (q1) at (3.5,-0.8);
\coordinate (q2) at (3.5,-2.1);
\coordinate (s)  at (3.5,-3.5);

\coordinate (c) at (6.5,0);


\foreach \u in {a1,a2,b1,b2} {
    \foreach \v in {p1,p2,q1,q2} {
        \draw[edge] (\u)--(\v);
    }
}

\foreach \u in {a1,a2,b1,b2} {
    \draw[edge] (t)--(\u);
}

\foreach \v in {p1,p2,q1,q2} {
    \draw[edge] (c)--(\v);
}

\foreach \v in {p1,p2} {
    \draw[edge] (d)--(\v);
}

\foreach \v in {q1,q2} {
    \draw[edge] (e)--(\v);
}

\foreach \u in {a1,a2} {
    \draw[edge] (r)--(\u);
}

\foreach \u in {b1,b2} {
    \draw[edge] (s)--(\u);
}


\node[vertex] (T) at (t) {$5$};

\node[vertex] (D)  at (d)  {$2$};
\node[vertex] (A1) at (a1) {$1$};
\node[vertex] (A2) at (a2) {$2$};
\node[vertex] (B1) at (b1) {$2$};
\node[vertex] (B2) at (b2) {$3$};
\node[vertex] (E)  at (e)  {$2$};

\node[vertex] (R)  at (r)  {$5$};
\node[vertex] (P1) at (p1) {$4$};
\node[vertex] (P2) at (p2) {$5$};
\node[vertex] (Q1) at (q1) {$5$};
\node[vertex] (Q2) at (q2) {$6$};
\node[vertex] (S)  at (s)  {$5$};

\node[vertex] (C) at (c) {$2$};


\node[lab,left=4pt] at (T.west) {$t$};

\node[lab,left=4pt] at (D.west)  {$d$};
\node[lab,left=4pt] at (A1.west) {$a_1$};
\node[lab,left=4pt] at (A2.west) {$a_2$};
\node[lab,left=4pt] at (B1.west) {$b_1$};
\node[lab,left=4pt] at (B2.west) {$b_2$};
\node[lab,left=4pt] at (E.west)  {$e$};

\node[lab,right=4pt] at (R.east)  {$r$};
\node[lab,right=4pt] at (P1.east) {$p_1$};
\node[lab,right=4pt] at (P2.east) {$p_2$};
\node[lab,right=4pt] at (Q1.east) {$q_1$};
\node[lab,right=4pt] at (Q2.east) {$q_2$};
\node[lab,right=4pt] at (S.east)  {$s$};

\node[lab,right=4pt] at (C.east) {$c$};

\end{tikzpicture}
\caption{A connected bipartite distance-hereditary graph
$G$ with $\chi_o(G)=6$. Consequently, a connected bipartite circle graph with odd chromatic number 6. The numbers inside the
vertices represent color of respective vertex.}
\label{fig:BDH6}
\end{figure}
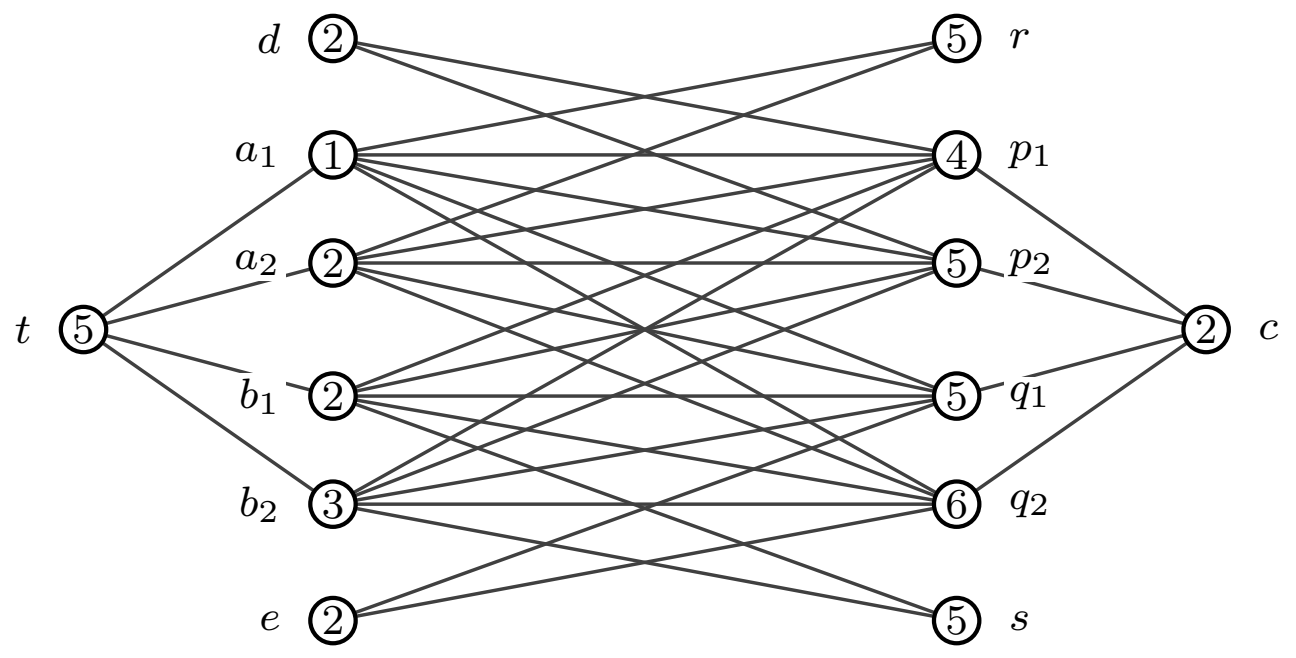
Let $G$ be a graph with bi-partition $X=\{a_1,a_2,b_1,b_2,c,d,e\}$ and $Y=\{p_1,p_2,q_1,q_2,r,s,t\}.$ Put $A=\{a_1,a_2,b_1,b_2\},
B=\{p_1,p_2,q_1,q_2\}.$ 
We show that $G$ is a connected bipartite
distance-hereditary graph and that
$\chi_o(G)=6$.

\medskip
\noindent\textbf{Claim 1.}
$G$ is a connected bipartite
distance-hereditary graph.

\smallskip
\begin{proof}
Clearly, $G$ is bipartite and connected. Recall that a connected bipartite graph is distance-hereditary if it can be constructed from a single vertex by successively adding pendant vertices and false twins.
We construct $G$ as follows.
Start with the edge $b_2t$ and perform the
following operations in the given order:
\begin{itemize}
    \item Add $a_2$ as a false twin of $b_2$.
    \item Add $r$ as a pendant vertex adjacent to $a_2$.
    \item Add $s$ as a pendant vertex adjacent to $b_2$.
    \item Add $q_2$ as a false twin of $t$.
    \item Add $c$ as a pendant vertex adjacent to $q_2$.
    \item Add $p_2$ as a false twin of $q_2$.
    \item Add $e$ as a pendant vertex adjacent to $q_2$.
    \item Add $q_1$ as a false twin of $q_2$.
    \item Add $d$ as a pendant vertex adjacent to $p_2$.
    \item Add $p_1$ as a false twin of $p_2$.
    \item Add $a_1$ as a false twin of $a_2$.
    \item Add $b_1$ as a false twin of $b_2$.
\end{itemize}

Each operation is performed on the graph
constructed so far. The resulting graph
is precisely $G$. Since pendant-vertex
and false-twin additions preserve
bipartite distance-hereditariness,
$G$ is a connected bipartite
distance-hereditary graph.
\end{proof}

\medskip
\noindent\textbf{Claim 2.}
$\chi_o(G)= 6$.


By \Cref{cor:cbdhg6}, we have $\chi_o(G) \leq 6$ for every connected bipartite distance-hereditary graph $G$. To establish equality, it remains to show that $\chi_o(G) \geq 6$.

Suppose, to the contrary, that $G$ admits
a proper odd coloring $\varphi$ using at
most five colors. Since $N(r)=\{a_1,a_2\}$ and $d_G(r)=2$, the odd coloring condition
implies $\varphi(a_1)\neq\varphi(a_2).$ Similarly, since $N(s)=\{b_1,b_2\},$ we obtain $\varphi(b_1)\neq\varphi(b_2).$ We first show that at least three distinct
colors are required on $A$.

Suppose that $A$ receives at most two
colors, say $\alpha$ and $\beta$.
The preceding inequalities imply $\{\varphi(a_1),\varphi(a_2)\}
=\{\alpha,\beta\}$ and $\{\varphi(b_1),\varphi(b_2)\}
=\{\alpha,\beta\}.$ However, $N(t)=\{a_1,a_2,b_1,b_2\}=A.$ Consequently, both $\alpha$ and $\beta$
appear exactly twice in $N(t)$.
Thus no color appears an odd number
of times in $N(t)$, contradicting
the odd coloring condition. Therefore, $|\varphi(A)|\geq 3,$ where $\varphi(A)$ denotes the set of
distinct colors used on $A$.

Similarly, since $N(d)=\{p_1,p_2\}, 
N(e)=\{q_1,q_2\},$ the odd coloring condition implies $\varphi(p_1)\neq\varphi(p_2), \varphi(q_1)\neq\varphi(q_2).$ Suppose that $B$ receives at most two
colors, say $\gamma$ and $\delta$.
Then $\{\varphi(p_1),\varphi(p_2)\}
=\{\gamma,\delta\}$ and $\{\varphi(q_1),\varphi(q_2)\}
=\{\gamma,\delta\}.$ Since $N(c)=\{p_1,p_2,q_1,q_2\}=B,$ both $\gamma$ and $\delta$ appear
exactly twice in $N(c)$.
This again contradicts the odd
coloring condition. Hence, $|\varphi(B)|\geq 3.$ Since every vertex of $A$ is adjacent
to every vertex of $B$, and $\varphi$
is proper, no color used on $A$ can
be used on $B$. Thus, $\varphi(A)\cap\varphi(B)=\varnothing.$ Consequently, the number of colors
used by $\varphi$ is at least
\[
\begin{aligned}
|\varphi(A)\cup\varphi(B)|
&=|\varphi(A)|+|\varphi(B)|\\
&\geq 3+3\\
&=6.
\end{aligned}
\]
This contradicts the assumption that
$\varphi$ uses at most five colors. Thus, $\chi_o(G)\geq 6.$ Therefore, $\chi_o(G)=6.$ 

Finally, since every bipartite distance-hereditary
graph is a bipartite circle graph, $G$ is also
a bipartite circle graph with odd
chromatic number $6$.

This completes the proof.

\end{proof}


\section{Biconvex Bipartite Graphs}
A bipartite graph $G=(X,Y,E)$ is \emph{biconvex} if there exist linear orderings of $X$ and $Y$ such that, for every vertex $v\in X\cup Y$, the neighbors of $v$ occur consecutively in the ordering of the partite set opposite to the one containing $v$. A bipartite graph $G=(X,Y,E)$ is a \emph{bipartite permutation graph} if and only if it admits a \emph{strong ordering}. A strong ordering consists of linear orderings $<_{X}$ of $X$ and $<_{Y}$ of $Y$ such that, for all $x,~x'\in X$ and $y,~y'\in Y$, if $xy,~x'y'\in E(G)$, $x<_{X}x'$, and $y<_{Y}y'$, then $xy'\in E(G)~\text{and}~x'y\in E(G)$.
\begin{definition}[Multichain Ordering \cite{biconvex_multichain_1,biconvex_multichain_2}]\label{def:mc}
Given a connected graph $G = (V,E)$, there exists a vertex $v_0 \in V (G)$ from which we construct the distance layers $L_0, L_1,\ldots, L_p$. The layer $L_i$, where $i \in [p]$, represents the set of vertices that are at distance $i$ from $v_0$. Here, $p$ denotes the largest integer such that $L_p$ is non-empty.
These layers form a \emph{multi-chain ordering} of $G$ if for every two consecutive layers $L_i$ and $L_{i+1}$, where $i \in \{0,1,\ldots,p-1 \}$, we have the edges connecting these layers form a chain graph. 
\end{definition}

\begin{theorem}
\label{thm:biconvexgraph}
    Let $G=(V,E)$ be a connected biconvex graph, $2 \le \chi_o(G) \le 5$.
\end{theorem}
\begin{proof}
    Since $G$ is connected and has at least two vertices, $\chi(G)\geq 2$. Moreover, every odd coloring is a proper coloring, and hence, $\chi_o(G)\geq \chi(G)\geq 2$. It remains to prove that $\chi_o(G)\leq 5$. Since $G$ is biconvex, it admits a multi-chain ordering $L_0,L_1,\ldots,L_p$, where $|L_0|=1$ and $L_i$ consists of the vertices at distance $i$ from the unique vertex in $L_0$. Note that, each layer $L_{i}$ is an independent set in $G$. Moreover, there are no edges between vertices if any two nonconsecutive layers and for each $i \in [p]$, $G[L_{i-1} \cup L_{i}]$ is a chain graph.
    Consequently, for each $i \in [p]$, there exists a vertex in $L_{i-1}$ that is adjacent to all vertices of $L_{i}$. We denote this vertex of $L_{i-1}$ as $v_{i-1}$ and call it {\em special vertex} in the layer $L_{i-1}$. Since $L_p$ is the final nonempty layer and there is no subsequent layer to dominate, we choose an arbitrary vertex $v_p\in L_p$ and designate it as the {\em special vertex} of $L_p$. Let $S =\{v_i \mid v_i \text{ is special vertex of $L_{i}$}\}$ be the set of special vertices. Next, we define a coloring function $f$ that uses at most $5$-colors, namely $\{0, 1,2,3,4\}$ and $f$ is an odd coloring of $G$. To do so, we first define colors on the set of special vertices $S$.

\[
f(v_{i}) = \begin{cases}
            0, & i \equiv 0 \, (\modo \, 3) \2, \\
            1, & i \equiv 1 \, (\modo \, 3) \2 ,\\
            2, & i \equiv 2 \, (\modo \, 3) \2 .\\
		 \end{cases}
\]

Now we define colors on the non-special vertices of each layer $L_{i}$ depending on whether $i$ is even or odd. For a each vertex $v \in L_{i} \setminus \{v_i\}$ we define the color on $v$ as follows:
    

\[
f(v)=
\begin{cases}
3,&i\text{ is even},\\
4,&i\text{ is odd}.
\end{cases}
\]

Consequently, each non special vertex of layer $L_{i}$ is colored $3$ if $i$ is even, 4 otherwise. Also special vertex of layer $L_{i}$ receives a color from the set $\{0,1,2\}$. An illustration of our colouring procedure is given in Fig.~\ref{fig:biconvex-multichain}.

\begin{figure}[ht]
\centering

\resizebox{0.80\textwidth}{!}{%
\begin{tikzpicture}[
    scale=0.40,
    transform shape,
    vertex/.style={
        circle,
        draw=black,
        fill=white,
        line width=0.3pt,
        minimum size=3mm,
        inner sep=0pt,
        font=\scriptsize
    },
    layer/.style={
        ellipse,
        draw=black,
        minimum width=0.6cm,
        minimum height=3cm
    },
    every label/.style={
        font=\scriptsize,
        text=black
    }
]


\node[layer] at (0,0) {};
\node[layer] at (2.4,0) {};
\node[layer] at (4.8,0) {};
\node[layer] at (7.2,0) {};
\node[layer] at (9.6,0) {};
\node[layer] at (12,0) {};

\node at (0,2.1) {$L_0$};
\node at (2.4,2.1) {$L_1$};
\node at (4.8,2.1) {$L_2$};
\node at (7.2,2.1) {$L_3$};
\node at (9.6,2.1) {$L_4$};
\node at (12,2.1) {$L_p$};


\node[vertex,label=below:{$v_0$}]
(v0) at (0,0) {0};


\node[vertex,label=below left:{$v_1$}]
(v1) at (2.4,-1.15) {1};

\node[vertex,label=left:{}]
(u11) at (2.4,-0.35) {4};

\node[vertex,label=left:{}]
(u12) at (2.4,0.35) {4};

\node[vertex,label=left:{}]
(u13) at (2.4,0.90) {4};


\node[vertex,label=below left:{$v_2$}]
(v2) at (4.8,-1.25) {2};

\node[vertex,label=right:{}]
(u21) at (4.8,-0.45) {3};

\node[vertex,label=right:{}]
(u22) at (4.8,0.05) {3};

\node[vertex,label=right:{}]
(u23) at (4.8,0.55) {3};

\node[vertex,label=right:{}]
(u24) at (4.8,0.95) {3};


\node[vertex,label=below left:{$v_3$}]
(v3) at (7.2,-1.25) {0};

\node[vertex,label=right:{}]
(u31) at (7.2,-0.45) {4};

\node[vertex,label=right:{}]
(u32) at (7.2,0.05) {4};

\node[vertex,label=right:{}]
(u33) at (7.2,0.55) {4};

\node[vertex,label=right:{}]
(u34) at (7.2,0.95) {4};


\node[vertex,label=below left:{$v_4$}]
(v4) at (9.6,-1.15) {1};

\node[vertex,label=left:{}]
(u41) at (9.6,-0.35) {3};

\node[vertex,label=left:{}]
(u42) at (9.6,0.35) {3};

\node[vertex,label=left:{}]
(u43) at (9.6,0.90) {3};


\node[vertex,label=below left:{$v_p$}]
(vp) at (12,-1.15) {2};

\node[vertex,label=right:{}]
(up1) at (12,-0.35) {4};

\node[vertex,label=right:{}]
(up2) at (12,0.35) {4};

\node[vertex,label=right:{}]
(up3) at (12,0.90) {4};


\draw (v0) -- (v1);
\draw (v0) -- (u11);
\draw (v0) -- (u12);
\draw (v0) -- (u13);

\draw (u13) -- (u24);
\draw (u12) -- (u24);
\draw (u11) -- (u24);

\draw (v1) -- (v2);
\draw (v1) -- (u21);
\draw (v1) -- (u22);
\draw (v1) -- (u23);
\draw (v1) -- (u24);

\draw (u21) -- (u32);
\draw (u21) -- (u31);
\draw (u21) -- (u34);
\draw (u21) -- (u33);
\draw (u21) -- (v3);

\draw (v2) -- (v3);
\draw (v2) -- (u31);
\draw (v2) -- (u32);
\draw (v2) -- (u33);
\draw (v2) -- (u34);

\draw (v3) -- (v4);
\draw (v3) -- (u41);
\draw (v3) -- (u42);
\draw (v3) -- (u43);

\draw (u41) -- (up3);
\draw (u42) -- (up3);
\draw (u43) -- (up3);

\draw (v4) -- (vp);
\draw (v4) -- (up1);
\draw (v4) -- (up2);
\draw (v4) -- (up3);

\end{tikzpicture}%
}

\caption{A layered graph with ordering
$L_0,L_1,\ldots,L_p$ for $p=5$
and special vertices $v_0,v_1,\ldots,v_p$.
The numbers inside the vertices indicate
assigned color labels, while vertex
names appear outside the vertices.}

\label{fig:biconvex-multichain}

\end{figure}
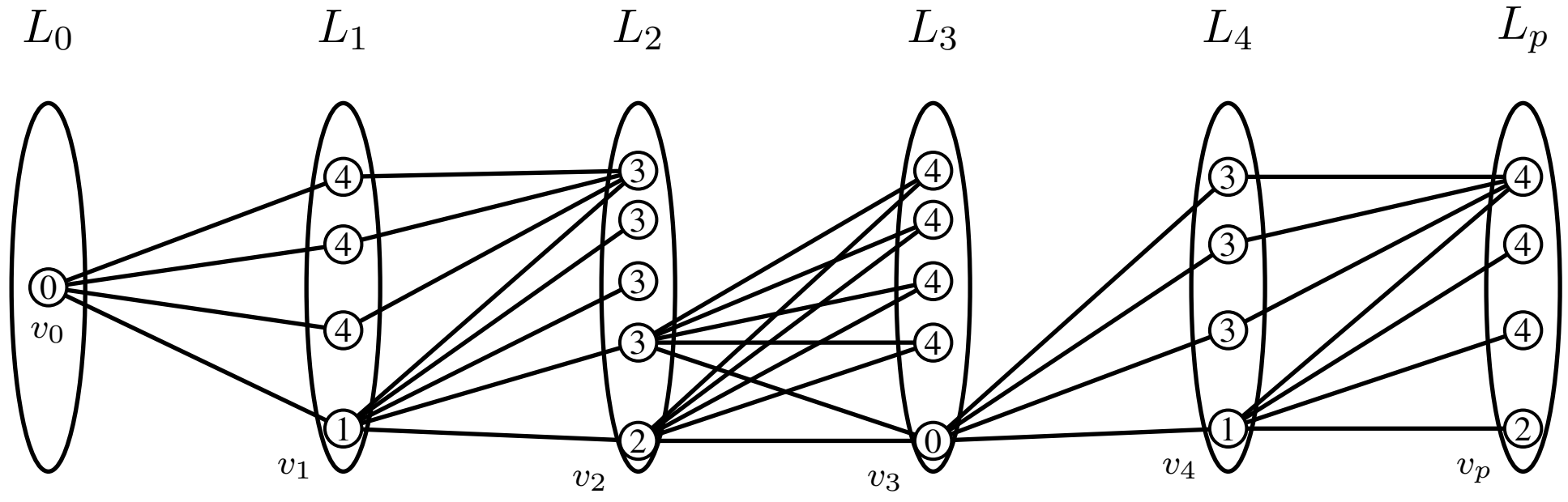

Next, we claim that $f$ is an odd coloring of $G$. In order to prove this, we first show that $f$ is a proper coloring in $G$. Let $v \in V(G)$ be any arbitrary vertex such that $v \in L_{i}$. Due to the property of multi-chain ordering, we note that $N(v) \subseteq L_{i-1} \cup L_{i+1}$. We analyze the colors on $N[v]$ by considering the following cases:

\begin{itemize}
    \item If $i$ is even.

    In this case if $v$ is a non-special vertex then $f(v)=3$. Moreover, if $v=v_i$ then $f(v) \in \{0,1,2\}$. Since both $i-1$ and $i+1$ are odd, for each non-special vertex $u \in L_{i-1} \cup L_{i+1}$, $f(u) =4$. Moreover, due to our coloring procedure, if $f(v_i) \equiv j \, (\modo \, 3) $ then $f(v_{i-1}) \equiv (j-1) \, (\modo \, 3)$ and $f(v_{i+1}) \equiv (j+1) \, (\modo \, 3) $. But for any three consecutive integers $m-1$, $m$, and $m+1$, the values $(m-1) \, \modo \, 3$, $(m) \, \modo \, 3$, and $(m+1) \, \modo \, 3$, all are distinct values from the set $\{0,1,2\}$. Consequently, $f(v_i) \neq f(v_{i-1})\neq f(v_{i+1})$. This concludes that $f(v) \neq f(u)$ for any $u \in N_{G}(v)$.

    \item If $i$ is odd.

    This is similar to the case above.
   
\end{itemize}

Thus, $f$ is a proper coloring of $G$. Next, we show that for each vertex $v \in V(G)$ there exist a color $c \in \{0,1,2,3,4\}$ such that $|f^{-1}(c) \cap N_{G}(v)|$ is odd. 
Consider vertices of any layer $L_{i}$, $i\ge1$.  Note that $v_{i-1}$ is adjacent to all vertices of the layer $L_{i}$. As discussed above, for any vertex $v \in L_{i}$, $N(v) \subseteq L_{i-1} \cup L_{i+1}$. Moreover if $f(v_{i-1}) =j$ then, $f^{-1}(j)  \cap (L_{i} \cup L_{i+1}) = \emptyset$. Also, as per our coloring procedure, $f^{-1}(j)  \cap L_{i-1}  = \{v_{i-1}\}$. Hence, for each vertex $v \in L_{i}$, $|f^{-1}(j)  \cap N(v)| = 1$.

Now, consider the layer $L_{0} = \{v_{0}\}$. Note that $N(v_0) \subseteq L_1$. In fact, $v_0$ is adjacent to all vertices of $L_{1}$. Let $f(v_1)=j$. As per our coloring procedure, $f^{-1}(j)  \cap L_{1}  = \{v_{1}\}$. Thus, $|f^{-1}(j)  \cap N(v_0)| = 1$. This concludes that $f$ is an odd coloring of $G$ that uses at most $5$ colors. Thus, the result follows.


\end{proof}

\begin{corollary}
The bounds $2\leq \chi_o(G)\leq 5$ for connected biconvex graphs $G$ with at least two vertices are sharp.
\end{corollary}

\begin{proof}
The lower bound is attained by the complete graph $K_2$, which is biconvex graph and satisfies $\chi_o(K_2)=2$. The upper bound is attained by the connected biconvex graph constructed in the following Proposition~\ref{pro:chi=5}, for which $\chi_o(G)=5$. Hence, both bounds are sharp.
\end{proof}

\begin{proposition}
\label{pro:chi=5}
There exists a connected biconvex bipartite graph $G$ such that $\chi_o(G)=5$.
\end{proposition}

\begin{figure}[htbp]
\centering
\begin{tikzpicture}[
    x=0.90cm, y=0.75cm,
    dot/.style={circle, draw=black, fill=black,
                minimum size=1mm, inner sep=0pt},
    cv/.style={circle, draw=black, thick,
               minimum size=3.5mm, inner sep=0pt,
               font=\small},
    edge/.style={black, semithick},
    col1/.style={fill=yellow!70},
    col2/.style={fill=cyan!45},
    col3/.style={fill=green!50},
    col4/.style={fill=orange!60},
    col5/.style={fill=red!55}
]

\def\drawGedges{
    \draw[edge] (0,4)--(3.1,4) (0,4)--(3.1,3);
    \draw[edge] (0,3)--(3.1,4) (0,3)--(3.1,3)
                (0,3)--(3.1,2) (0,3)--(3.1,1);
    \draw[edge] (0,2)--(3.1,4) (0,2)--(3.1,3)
                (0,2)--(3.1,2) (0,2)--(3.1,1)
                (0,2)--(3.1,0);
    \draw[edge] (0,1)--(3.1,4) (0,1)--(3.1,3)
                (0,1)--(3.1,2) (0,1)--(3.1,1)
                (0,1)--(3.1,0);
    \draw[edge] (0,0)--(3.1,2) (0,0)--(3.1,1);
}

\begin{scope}
\node at (1.55,5.3) {Original graph};
\node at (0,4.7) {$X$};
\node at (3.1,4.7) {$Y$};

\drawGedges

\node[dot,label=left:$a$] at (0,4) {};
\node[dot,label=left:$c$] at (0,3) {};
\node[dot,label=left:$p$] at (0,2) {};
\node[dot,label=left:$q$] at (0,1) {};
\node[dot,label=left:$b$] at (0,0) {};

\node[dot,label=right:$y_1$] at (3.1,4) {};
\node[dot,label=right:$y_2$] at (3.1,3) {};
\node[dot,label=right:$y_3$] at (3.1,2) {};
\node[dot,label=right:$y_4$] at (3.1,1) {};
\node[dot,label=right:$y_5$] at (3.1,0) {};

\end{scope}

\begin{scope}[xshift=5.8cm]
\node at (1.55,5.3) {Odd 5-coloring};
\node at (0,4.7) {$X$};
\node at (3.1,4.7) {$Y$};

\drawGedges

\node[cv,label=left:$a$] at (0,4) {2};
\node[cv,label=left:$c$] at (0,3) {2};
\node[cv,label=left:$p$] at (0,2) {1};
\node[cv,label=left:$q$] at (0,1) {2};
\node[cv,label=left:$b$] at (0,0) {5};

\node[cv,label=right:$y_1$] at (3.1,4) {3};
\node[cv,label=right:$y_2$] at (3.1,3) {5};
\node[cv,label=right:$y_3$] at (3.1,2) {3};
\node[cv,label=right:$y_4$] at (3.1,1) {4};
\node[cv,label=right:$y_5$] at (3.1,0) {4};

\end{scope}

\end{tikzpicture}
\caption{A connected biconvex bipartite graph $G$ and an odd $5$-coloring of $G$. The numbers inside the vertices
represent color of respective vertex.}
\label{fig:biconvex-two-side}
\end{figure}
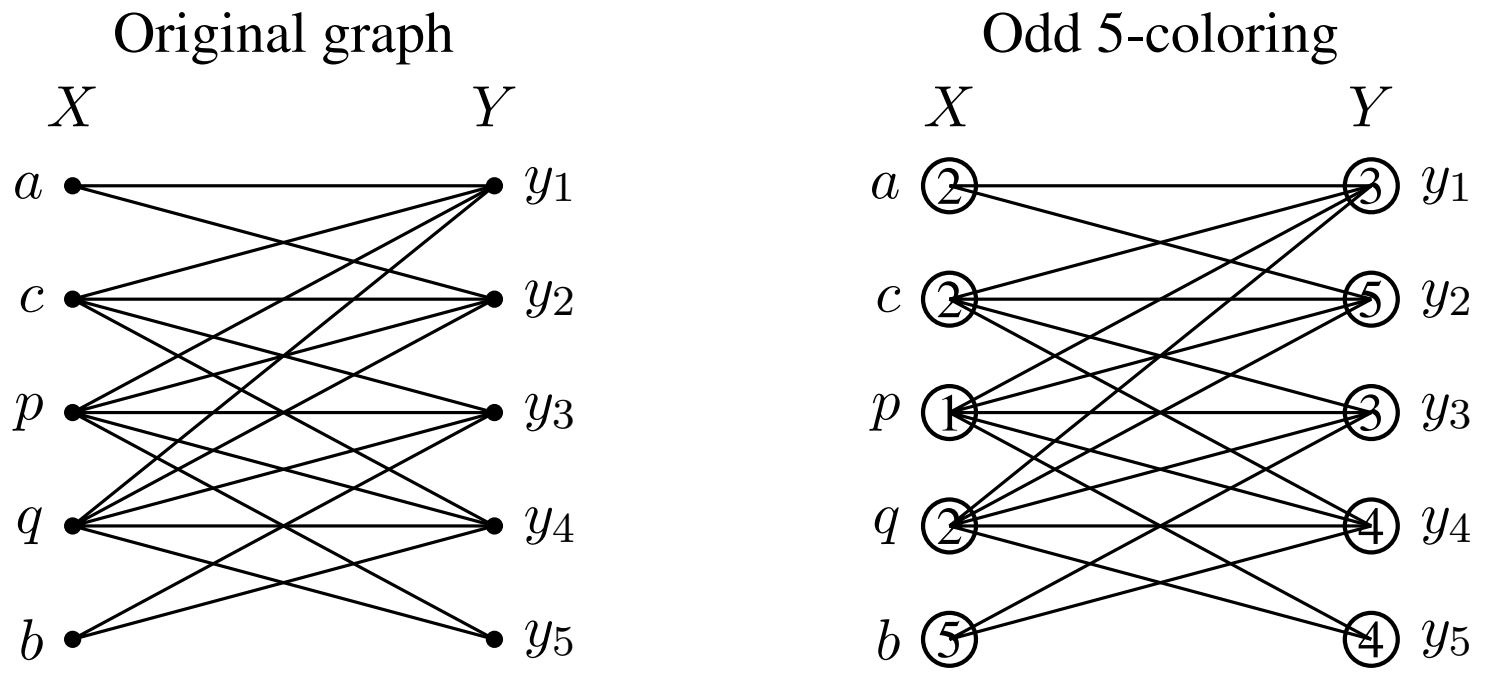

\begin{proof}
Consider the bipartite graph $G=(X,Y,E)$ illustrated in Fig.~\ref{fig:biconvex-two-side}, where $X=\{a,c,p,q,b\}
\text{ and } Y=\{y_1,y_2,y_3,y_4,y_5\}$. Clearly, the neighbors of every vertex of $X$ occur consecutively in the ordering of $Y$. Similarly the neighbors of every vertex of $Y$ also occur consecutively in the ordering of $X$. Consequently, $G$ is biconvex. Furthermore, 
the graph $G$ is connected.



By \Cref{thm:biconvexgraph}, we have $\chi_o(G) \leq 5$ for every connected biconvex graph $G$. To establish equality, it remains to show that $\chi_o(G) \geq 5$.



Suppose, to the contrary, that $G$ admits an odd coloring $f$ using at most four colors. Since $N(y_5)=\{p,q\}$, the odd-coloring condition at $y_5$ implies that $f(p)\neq f(q).$ Without loss of generality, assume $f(p)=1 \text{ and }
f(q)=2.$ Since $p$ and $q$ are adjacent to every vertex of $Y$, and $f$ is a proper coloring, no vertex of $Y$ can receive
color $1$ or color $2$. Therefore, every vertex of $Y$ must receive one of the remaining two colors, say $3$ and $4$.

Now, consider the vertex $a$. Since $N(a)=\{y_1,y_2\},$ the odd-coloring condition at $a$ implies that $f(y_1)\neq f(y_2).$ Hence, colors $3$ and $4$ each occur exactly once in $\{y_1,y_2\}$. Similarly, since $N(b)=\{y_3,y_4\},$ the odd-coloring condition at $b$ implies that $f(y_3)\neq f(y_4).$ Thus, colors $3$ and $4$ each occur exactly once in $\{y_3,y_4\}$.

Finally, consider the vertex $c$. We have $N(c)=\{y_1,y_2,y_3,y_4\}.$ By the above observations, each of
the colors $3$ and $4$ occurs exactly twice in $N(c)$, while every other color occurs zero times. Therefore, no color occurs an odd number of times in $N(c)$. This contradicts the assumption that $f$ is an odd coloring of $G$. Consequently, $\chi_o(G)\geq 5.$



This completes the proof.
\end{proof}


\section{Cographs}

In this section, we study the odd chromatic number of a $C_4$-free cographs. A graph $G$ is a \emph{cograph} if it contains no induced subgraph isomorphic to $P_4$. A \emph{$C_4$-free cograph} is a cograph that contains no induced subgraph isomorphic to $C_4$. The odd chromatic number of a cograph is already known.

\begin{theorem}{\rm \cite{bhyravarapu2025parameterized}}
Let $G$ be a cograph. Then $\omega(G) \le \chi_o(G)\leq \omega(G)+2$.
\end{theorem}

In fact, we observe that the above-proposed bound is tight. We reduced this bound for a cograph in the absence of $C_4$. Recall that a $\{C_4,P_4\}$-free graph is also known as {\em trivially perfect graph}~\cite{Golumbic2022,golumbic1978trivially}. A $C_4$-free cograph $G$ can be obtained from $K_1$ by repeatedly applying one of the following in each step:
\begin{enumerate}
   \item add an isolated vertex,
   \item add a universal vertex,
   \item take the disjoint union of two $C_4$-free cographs $G_1$ and $G_2$.
\end{enumerate}

Next, we state the following theorem:

\begin{theorem}
\label{thm:cograph}
Let $G$ be a $C_4$-free cograph. Then $\omega(G)\leq\chi_o(G)\leq\omega(G)+1$.
\end{theorem}
\begin{proof}

Let $G$ be a $C_4$-free cograph. Every $C_4$-free cograph is perfect, and hence
$\chi(G)=\omega(G)$. Since every odd coloring is a proper coloring, $\omega(G)= \chi(G)\leq \chi_o(G)$. It remains to establish the upper bound. We proceed by induction on $|V(G)|$. 
For $G=K_1$, $\chi_o(G)=\omega(G)=1$. Now, let us assume that, for every $C_4$-free cograph $H$ with $|V(H)|<|V(G)|$ we have $\chi_o(H)\leq\omega(H)+1$. 

\begin{itemize}
    \item \textbf{Case 1: $G=H\cup\{v\}$.} 
    
    Let $G=H\cup\{v\}$, where $deg_G(v)=0$. Since $v$ has no neighbors, it imposes no odd-coloring constraint. Hence, $\chi_o(G)=\chi_o(H)$. Furthermore, $\omega(G)=\omega(H)$. Therefore, by induction, $\chi_o(G)=\chi_o(H)\leq\omega(H)+1=\omega(G)+1$.

    \medskip
    \item \textbf{Case 2: $G$ is disjoint union of two $C_4$-free cographs $G_1$ and $G_2$.}

    Let $G=G_1\cup G_2$. Since there are no edges between $G_1$ and $G_2$, then $\omega(G)=\max\{\omega(G_1),\omega(G_2)\}$ and
    $\chi_o(G)=\max \{\chi_o(G_1),\chi_o(G_2)\}$. By the induction hypothesis, $\chi_o(G_i) \leq \omega(G_i)+1$ for $i\in\{1,2\}$. Consequently, $\chi_o(G) \leq \max \{\omega(G_1)+1, \omega(G_2)+1\}= \omega(G)+1$.

    \item \textbf{Case 3: $G=\{v\}\vee H$.} 

    Let $G=\{v\}\vee H$, where $v$ is universal vertex, and put $k=\omega(H)$. Then $\omega(G)=k+1$. Since $H$ is perfect, $\chi(H)= \omega(H)=k$. Let $c:V(H)\rightarrow[k]$ be a proper $k$-coloring of $H$. Choose an arbitrary vertex $x\in V(H)$ and define $f:V(G)\rightarrow[k+2]$ by $f(v)=k+1$, $f(x)=k+2$, and $f(u)=c(u)$ for every $u\in V(H)\setminus\{x\}$.

    Clearly, $f$ is proper. Since $v$ is universal vertex and $x$ is the unique vertex assigned color $k+2$, $|N(v)\cap f^{-1}(k+2)|=1$. Moreover, $v$ is the unique vertex assigned color $k+1$, and hence $|N(u)\cap f^{-1}(k+1)|=1$ for every $u\in V(H)$. Thus,
    $\forall~ u\in V(G),~\exists~ i \in[k+2]$ such that $|N(u)\cap f^{-1} (i)| \equiv1 \pmod2$. Therefore, $f$ is an odd coloring of $G$, and $\chi_o(G)\leq k+2=\omega(G)+1$.

    Thus, the upper bound holds in each of the three construction steps. By induction,
    $\chi_o(G)\leq\omega(G)+1$. Combining this with the lower bound gives $\omega(G) \leq \chi_o(G)\leq\omega(G)+1$.
\end{itemize}

This completes the proof.

\end{proof}

\begin{corollary}
    The above proposed bounds are tight. For $G=K_n$, $\chi_o(G)=\omega(G)=n$, whereas for $G=P_3$, $\omega(G)=2$ and $\chi_o(G)=3=\omega(G)+1$. 
\end{corollary}

\section{Concluding Remarks}
In this paper, we established bounds on the odd chromatic number for several important graph classes and characterized these classes in terms of it. As a consequence, we also settle the conjecture by Pang et al. in \cite{pang2026odd} for block-cactus graphs, block graphs, proper interval graphs, biconvex bipartite graphs and $C_4$-free cographs. We answer Problem~5.2 posed by Davies et al. \cite{Davies2025OddColoring} in the negative by constructing a bipartite circle graph with odd chromatic number five. For a proper interval graph $G$, we proved that $\chi_{o}(G) \le \omega(G)+1$. A proper interval graph is a subclass of class of chordal graphs. It would be interesting to obtain bounds on the odd chromatic number of a chordal graph. Establishing bounds on the odd chromatic number of other important graph classes, such as distance-hereditary graphs, is also a promising research direction.




\bibliographystyle{plainurl}
\bibliography{Bibfile}

\end{document}